\documentclass[a4paper]{jpconf}
\usepackage{graphicx}
\usepackage[utf8]{inputenc}

\usepackage[T1]{fontenc}
\usepackage{cite}

\def\df{\partial}
\def\arcsinh{\mathop{\mbox{arcsinh}}}
\def\diag{\mathop{\mbox{diag}}}

\begin{document}

\title{Wormholes and Teleporters}

\author{Igor Nikitin}

\address{Fraunhofer Institute for Algorithms and Scientific Computing,\\
Schloss Birlinghoven, 53757 Sankt Augustin, Germany}

\ead{igor.nikitin@scai.fraunhofer.de}

\begin{abstract}
In general relativity, there is a class of solutions that currently do not have observed analogues, but on which the theory is shaped, giving an understanding what is fundamentally possible within its framework. Such solutions include wormholes, tunnels that connect distant regions in spacetime. Although not a single wormhole has yet been discovered, there is a large number of works devoted to their study, thanks to which wormholes as a class of solutions become firmly established in modern science. In this paper, we consider two topologically nontrivial types of solutions related to wormholes. First: wormholes that can open and close. In this relation, we will discuss topological censorship theorems, which under certain conditions prohibit changing topology. We will also discuss known ways to circumvent these theorems. Using analytical and numerical methods, as well as visualization, we will construct an example of an opening and closing wormhole with the dimensions of the central black hole in the Milky Way galaxy. Our construction continues the work by Kardashev, Novikov and Shatskiy, in which a static wormhole with the same parameters was considered. The second type is a modification of Visser's dihedral wormhole solution for a dynamic case, which can be considered as a model of a teleportation event.

\end{abstract}

\section{Introduction}\label{sec1}

This is the second part of the work started in arXiv:1909.08984, where quasistatic solutions of the opening wormholes type were introduced. Here we briefly review the results and consider their extension to the full dynamic case. We will also discuss the general question on changing topology in general relativity and consider in more detail a new type of solutions, topological teleporters. 

{\it Wormholes} are topologically non-trivial solutions of general theory of relativity, describing shortcuts or tunnels, connecting distant regions in spacetime. Their study starts with pioneering works by Einstein and Rosen in 1935, Wheeler 1955, through ``a renaissance'' of wormhole solutions by Morris and Thorn 1988, to recent works by Visser 1996 and Lobo 2016. The progress has been described in the book \cite{Visser1996}, the recent developments were reported in \cite{1604.02082}. There are many different types of wormholes, static and dynamic, micro- and macroscopic, traversable and non-traversable, with and without spherical symmetry, possessing various matter constitution. Relatively rare case, which will especially interest us, are the wormholes that can open and close. Their peculiarity is that these solutions possess {\it variable topology}. For general relativity, the possibility of the existence of such solutions has been considered in \cite{Visser1996}, on the basis of the original works \cite{Geroch1, Geroch2, Geroch-Horowitz, Borde, Hawking1, Hawking2, Hawking-Ellis}. On one hand, a number of so called topological censorship theorems have been formulated, that prohibit change of topology in a certain class of solutions. On the other hand, in works \cite{Yodzis1972,Sorkin1,Sorkin2,Horowitz1991,9109030,Morris-etal,9305009,Raju1982,CosmicCens,vickers1,9907105,9605060,9711069,0410087,0505150,conical_spacetimes} a different, slightly wider class has been identified, where the change of topology becomes possible. In the given paper, in Section~\ref{sec2}, we discuss what are exactly the consequences of topology change, the class of applicability of the topological censorship theorems and the structure of solutions in the extended class.

As an implementation of this general theory, in Section~\ref{sec3}, we will explicitly construct an example of {\it a dynamically opening wormhole}. For this purpose we use computer algebra and numerical integration methods, as well as visualization. The constructed solution possesses a specific topological signature: it has two copies of three-dimensional space in the initial state, while the final state contains a closed bubble (baby universe) and two copies of three-dimensional space connected by a wormhole. It also provides a milder singularity of matter distribution than the other proposals of this kind. The solution can be dimensioned to the sizes of the central black hole in the Milky Way galaxy. A similar scenario  with the static wormhole in the center of the Milky Way has been considered in paper \cite{0610441}.

{\it Teleportation} is a concept similar to wormholes, that also successfully paves the way from science fiction to the field of serious scientific research. In the past three decades, a large number of works have appeared on the so-called quantum teleportation, see report \cite{qtp2017} on recent advances. In this approach, quantum entangled states are used for a propagation of information about the quantum state of elementary particles located far from each other. At the same time, the classical general relativity also possesses solutions of this kind. A special position is taken by the concept coming from popular science that a portal or {\it stargate} can connect remote regions of the universe and can be used for traveling between them. Interestingly, a solution of this type was constructed by M.~Visser in 1989 \cite{Visser1,Visser2} and presented in his book \cite{Visser1996}. This is exact solution in general relativity that turned out to be a special type of wormhole, {\it dihedral wormhole}, from the wider class of polyhedral wormholes describing just such a portal. In Section~\ref{sec4}, we construct a modification of this solution by Wick rotation, a formal replacement of time with the complex coordinate $t\to iz$, and examine the relationship of the resulting solution with the other solutions from the wormholes class.

Two appendices provide the necessary technical details of the constructions.

\section{Topology change}\label{sec2}

General relativity considers spacetime as a manifold equipped with a Lorentzian metric $g_{\mu\nu}$ (a point-dependent 4x4 symmetric matrix of signature $(-,+,+,+)$). For coordinate variation $dx^\mu$ on the manifold, the length element is defined as a quadratic form $ds^2=g_{\mu\nu}dx^\mu dx^\nu$. The vectors with $ds^2<0$ are called timelike, with $ds^2>0$ -- spacelike, with $ds^2=0$ -- lightlike, or null. The inverse metric is denoted as $g^{\mu\nu}$ and used to raise and lower the indices, e.g., $V^\mu=g^{\mu\nu}V_\nu$, $V_\mu=g_{\mu\nu}V^\nu$, where the summation over repeating indices is standardly assumed. 

\begin{figure}
\begin{center}
\includegraphics[width=0.8\textwidth]{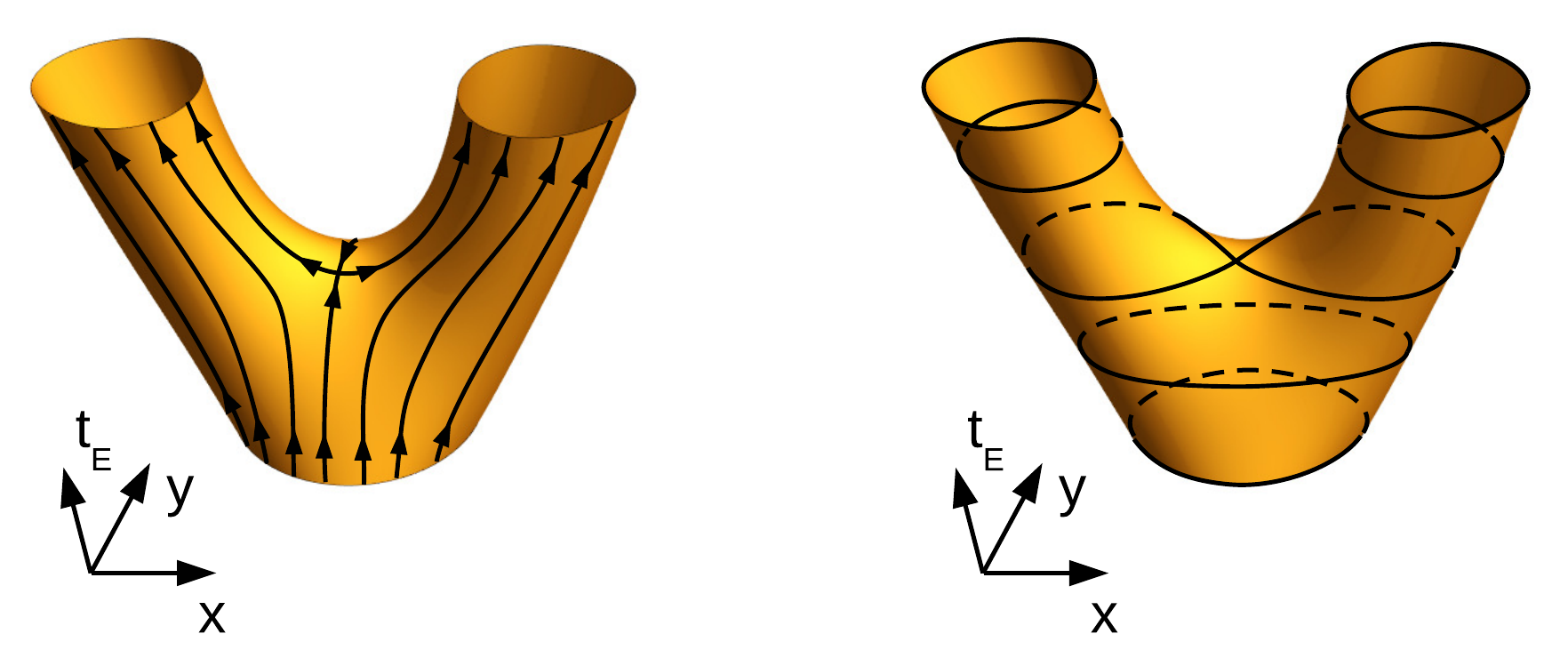}

\end{center}
\caption{Solutions of variable topology in general relativity. The process $ S ^ 1 \to2S ^ 1 $. On the left -- the vector field, defining the direction of time, possessing a singularity (saddle point). On the right -- equal-time slices, with corresponding Morse rearrangement.}\label{f3}
\end{figure}

\begin{figure}
\begin{center}
\includegraphics[width=0.8\textwidth]{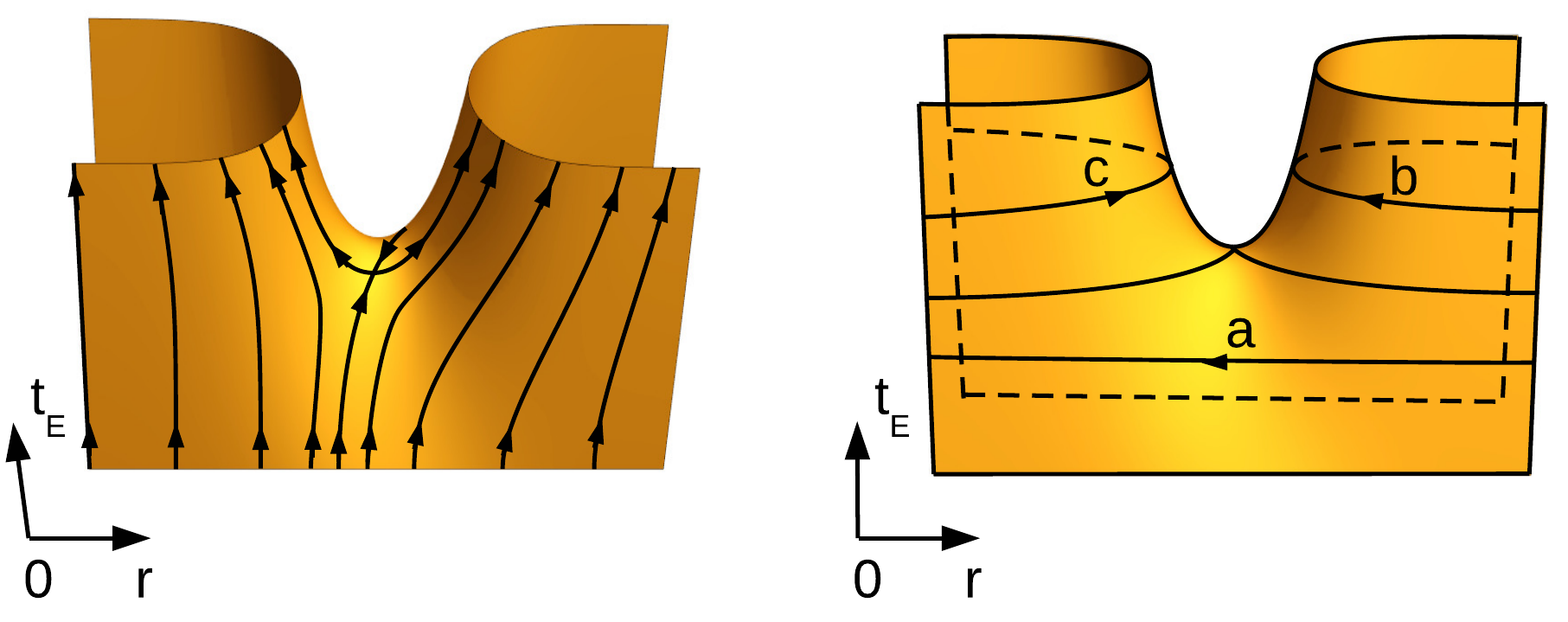}

\end{center}
\caption{The same plots for the process $ 2R ^ 3 \to R \times S ^ 2 + S ^ 3 $. The initially trivial slice (a) transforms to a wormhole (b) and a closed bubble (c).}\label{f4}
\end{figure}

\begin{figure}
\begin{center}
\includegraphics[width=0.8\textwidth]{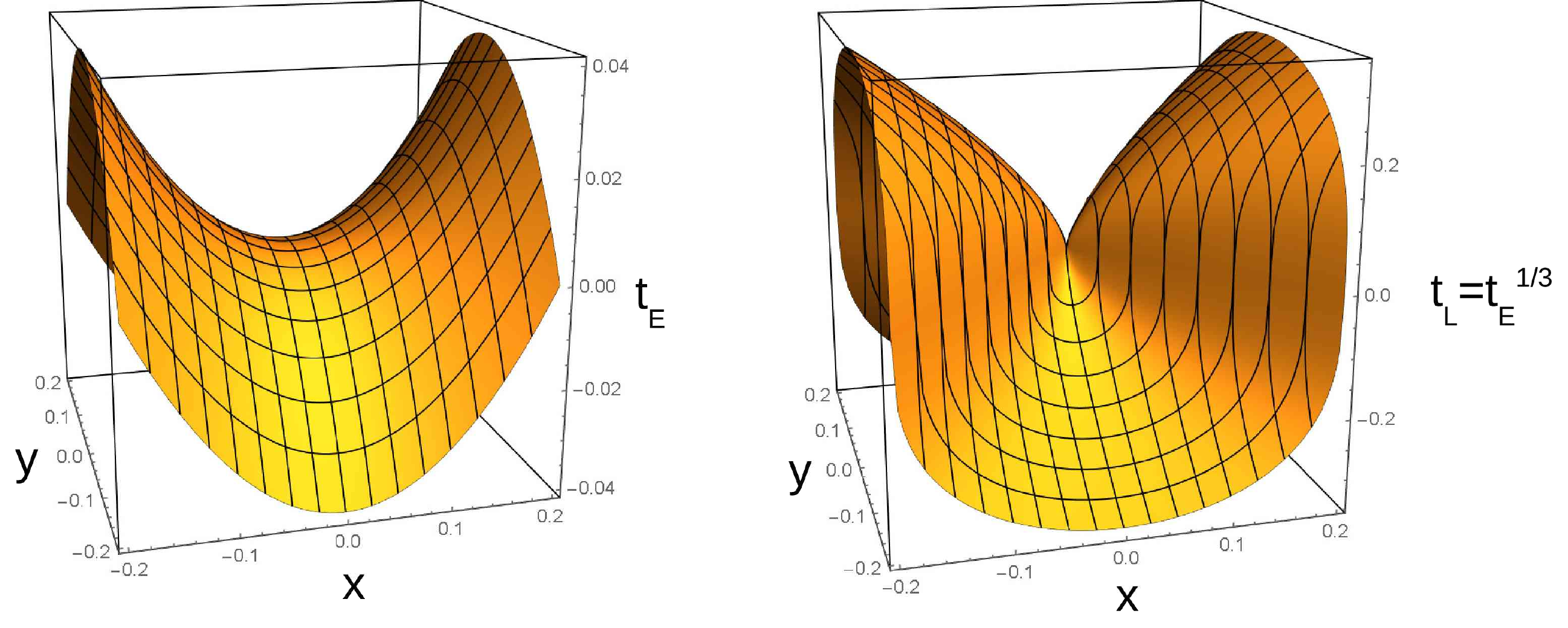}

\end{center}
\caption{The same process on the Euclidean embedding diagram (left) and the Lorentzian embedding diagram (right).}\label{f5}
\end{figure}

The topological type of spacetime manifold is initially not fixed. The possibility of topology change, or more precisely, the ability for spatial slices to change their topology over time, is one of disputed questions in general relativity. This question was examined in detail in \cite {Visser1996} Chap.6.5, based on the original works \cite {Geroch1, Geroch2, Geroch-Horowitz, Borde, Hawking1, Hawking2, Hawking-Ellis}. There is a number of so-called topological censorship theorems that prohibit topology change in special classes of spacetime manifolds, in particular, on {\it Lorentzian time-orientable chronological manifolds}. On such manifolds, in addition to the Lorentzian metric, there should be everywhere nonzero continuous vector field defining the direction of time; moreover, there should be no closed integral trajectories for this field (no closed timelike curves). On the other hand, \cite {Visser1996} also describes a way around these theorems. It involves the consideration of {\it almost everywhere Lorentzian manifolds}, in which the Lorentzian metric is introduced everywhere, except for a thin set of singular points. This approach has been chosen in papers \cite{Yodzis1972,Sorkin1,Sorkin2,Horowitz1991,9109030,Morris-etal,9305009,Raju1982,CosmicCens,vickers1,9907105,9605060,9711069}, the recent advances have been reported in \cite{0410087,0505150,conical_spacetimes}. At first, one notes, that the presence of singularities in general relativity is commonly accepted (well-known examples are pointlike Schwarzschild and ringlike Kerr singularities). The topology change simply adds a new type to the existing family of singularities. The proposed algorithm is to fix the topology of spacetime manifold, define the metric, calculate the Einstein tensor proportional to the energy-momentum tensor ($G_{\mu \nu}=8\pi T_{\mu \nu}$) and analyze the resulting singularities in the distribution of matter. To proceed along this way, there are three known methods for defining spacetime manifolds.

\begin{figure}
\begin{center}
\includegraphics[width=0.9\textwidth]{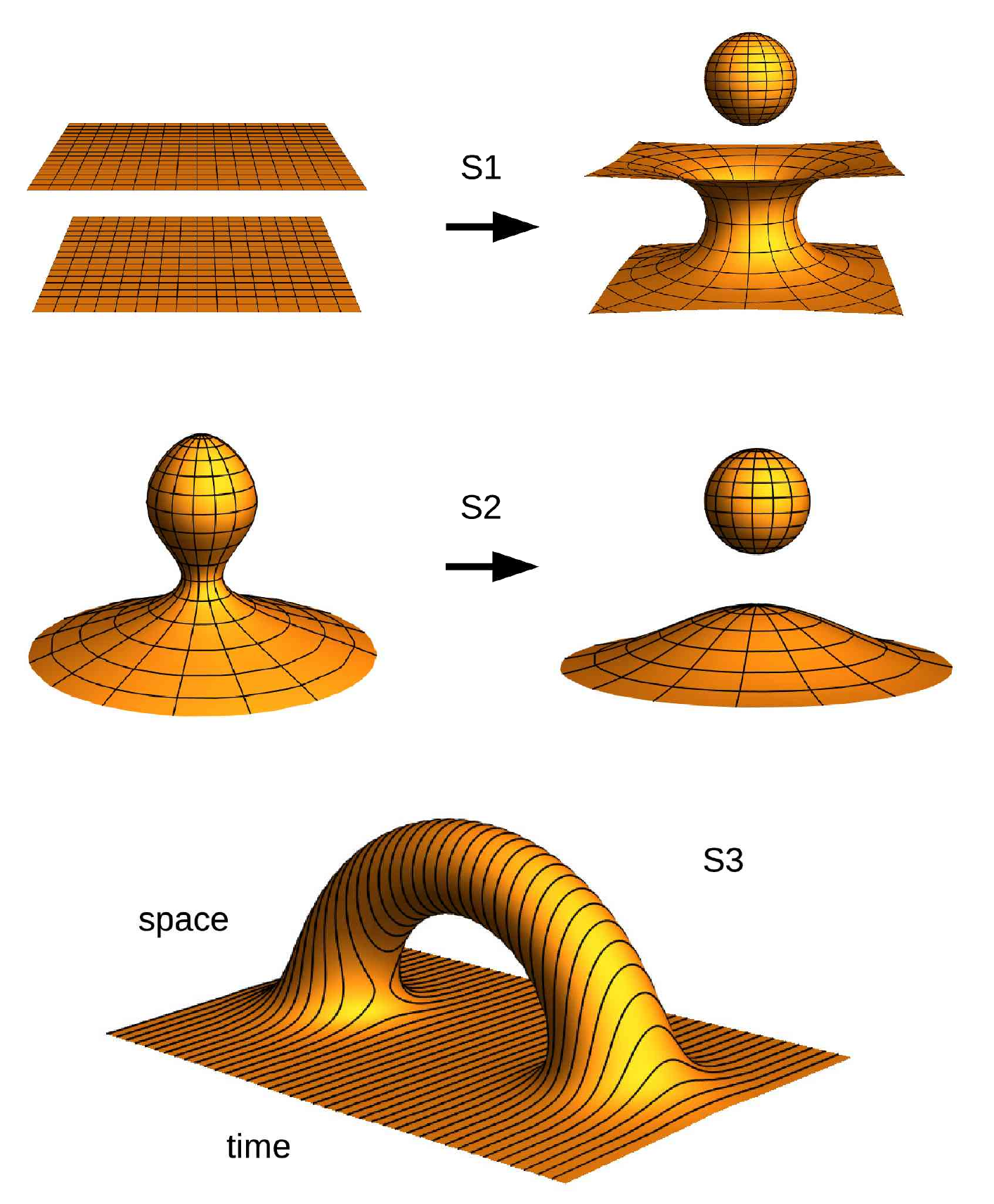}

\end{center}
\caption{Various scenarios of wormhole opening (see text).}\label{f7}
\end{figure}

\paragraph*{The first method} is to define a surface in Euclidean space on which the vector field $ V ^ \mu $ is specified. The metric of Euclidean space is induced on the surface and then redefined to the Lorentzian metric by the formula \cite{Visser1996}
\begin{equation}
(g_L)^{\mu\nu}=(g_E)^{\mu\nu}-2 V^\mu V^\nu /((g_E)_{\alpha\beta}V^\alpha V^\beta).\label{vecmetr}
\end{equation}
In fact, the metric is re-projected in the direction of the vector field, so that the components along it receive a Lorentzian signature. With respect to the new metric, the vector field $ V ^ \mu $ is timelike; it can be used to specify the direction of time on the manifold.

A particular choice of the vector field is $V=\df f$, a gradient of {\it Morse function} on the manifold. For topology change $M_0\to M_1$, an interpolating manifold, or {\it a cobordism} is a manifold $M$ whose boundary is a disjoint union of $M_0$ and $M_1$, $\df M= M_0\sqcup M_1$. Morse function is any smooth function, taking values $f(M_0)=a$, $f(M_1)=b$ on the boundaries and intermediate values in $M\backslash\df M$, whose {\it critical points} ($\df f=0$) are non-degenerate ($\det \df^2 f\neq0$) and located in $M\backslash\df M$. Morse function can be taken as a global time coordinate $t=f$, interpolating between the initial and final states in the topological transition. The Lorentzian metric then can be defined by a formula \cite{Sorkin2}:
\begin{equation}
(g_L)_{\mu\nu}=\delta_{\mu\nu}((\df_\alpha f)(\df_\beta f)\delta^{\alpha\beta})-\zeta(\df_\mu f)(\df_\nu f),\ \zeta>1.\label{morsemetr}
\end{equation}
Comparing it with the previous definition, we see, in addition to the replacement $V\to\df f$, (i) a trivial choice for the Euclidean metric $(g_E)_{\mu\nu}=\delta _{\mu\nu}$, (ii) an overall factor $((\df_\alpha f)(\df_\beta f)\delta^{\alpha\beta})$ and (iii) an arbitrary scaling factor $\zeta>1$ in the reprojection of the metric. These additional choices are optional and can be modified on necessity. 

In these definitions, the problems that may arise from changing the topology immediately become clear. Fig.\ref{f3} shows the process of splitting a closed universe into two closed universes in dimension 1, that is, $ S ^ 1 \to2S ^ 1 $. The vector field $ V ^ \mu $ is also shown (visible on one side, continued to the other side mirror symmetrically). It is easy to verify that a vector field on such manifold, with its direction fixed on boundary circles, necessarily has a singular point (the proof is based on Poincaré-Hopf theorem). Thus, a globally continuous nonzero vector field with the described boundary conditions does not exist for the manifold under consideration. Fig.\ref{f3} on the right shows the levels of Morse function, which necessarily has a critical point. In our case it is of saddle type, with a typical hyperbolic rearrangement of levels in its vicinity.

Fig.\ref{f4} shows the other solution with the similar structure. It shows the Euclidean embedding diagram for a spherically symmetric solution in 3-dimensional space, plus 1 time. The diagram defines the behavior of the metric for the radial and temporal components $ (r, t) $, while the remaining two angular coordinates have the standard spherical metric definition.

The solution describes {\it a dynamical opening of a wormhole}, accompanied by the separation of a bubble, a closed baby universe, with the topology $ 2R ^ 3 \to R \times S ^ 2 + S ^ 3 $. To verify this, consider the right side of the figure, which shows several time slices. The surface has two sheets, front and back, which correspond to two copies of spacetime. Line (a) corresponds to the path along the radius from large values to the center of the system, on one sheet of space. After the reconnection, curve (b) shows the path from one sheet to another, performed through the minimum value of $ r $, {\it wormhole throat}. Curve (c) connects two sheets through the maximum value of $ r $, the radius of the closed universe. The topological type $R \times S ^ 2$ of the wormhole is formed from the spherical throat $S ^ 2$ and a linear path through it: $R=(-\infty,\infty)$. The baby universe $S^3$ is formed from two balls $D^3$, glued along the boundary $S^2$. The process of a dynamical opening of a wormhole according to this scheme is the main topic that we will discuss further.

\paragraph*{The second method} is to construct embedding diagrams directly in the space of a Lorentzian signature. For example, a surface can be constructed in flat Minkowski spacetime, with metric $ ds ^ 2 = dt ^ 2-dx ^ 2-dy ^ 2 $ . After inducing the metric on the surface, one should make sure that it has a Lorentzian signature. There may be problems with this. Fig.\ref{f5} on the left shows the surface in the form of a hyperbolic paraboloid $ t_E = x ^ 2-y ^ 2 $. Induction of Minkowski metric onto it, obviously, will not lead to the Lorentzian metric, it fails to have the necessary signature in the vicinity of $x=y=0$. At the same time, if we apply the transformation $ t_L = t_E ^ {1/3} $, then we get {\it almost everywhere Lorentzian manifold}, of the same topology as the Euclidean diagram considered before. Straightforward computation shows that the induced metric is Lorentzian in a vicinity of $x=y=0$, except of this point, where the surface has a singularity.

\paragraph*{The third method} is a direct definition of the metric components, which for spherically symmetric solutions can be written in the standard way (see, e.g., \cite{Blau2018} Chap.23.6):
\begin{eqnarray}
&ds^2=-A(r,t)dt^2+B(r,t)dr^2+2C(r,t)dtdr+D(r,t)r^2(d\theta^2+\sin^2\theta\, d\phi^2). \label{stdmetr}
\end{eqnarray}
Further, after the metric is specified, the Einstein tensor can be evaluated with a straightforward algorithm, including a chain of substitutions, differentiations and algebraic simplifications. Practically, one can use the following Mathematica code \cite{math-codes}:

\vspace{3mm}\noindent{\bf Algorithm Einstein(n,x,g):}
{\footnotesize
\begin{verbatim}
     ginv = Simplify[Inverse[g]];
     gam = Simplify[ Table[
            (1/2) Sum[ 
               ginv[[i,s]] (D[g[[s,j]],x[[k]]]+D[g[[s,k]],x[[j]]]-D[g[[j,k]],x[[s]]]), 
            {s,1,n} ], 
          {i,1,n},{j,1,n},{k,1,n} ] ];
     R4 = Simplify[ Table[
            D[gam[[i,j,l]],x[[k]]]-D[gam[[i,j,k]],x[[l]]] 
            + Sum[ gam[[s,j,l]] gam[[i,k,s]] - gam[[s,j,k]] gam[[i,l,s]], {s,1,n} ],
          {i,1,n},{j,1,n},{k,1,n},{l,1,n} ] ];
     R2 = Simplify[ Table[
            Sum[ R4[[i,j,i,l]],{i,1,n} ], {j,1,n},{l,1,n} ] ];
     R0 = Simplify[ Sum[ ginv[[i,j]] R2[[i,j]], {i,1,n},{j,1,n} ] ];
     G2 = Simplify[ R2 - (1/2) R0 g ]
\end{verbatim}}
The algorithm takes as an input the metric $g_{\mu\nu}(x)$ with $n=\dim(x)$ and evaluates the Einstein tensor $G_{\mu\nu}(x)$, proportional to the energy-momentum tensor $T_{\mu \nu}=G_{\mu \nu}/(8\pi)$.

\vspace{3mm}
We prefer to use the representation (\ref{stdmetr}) to define solutions with dynamically changing topology. First, we consider the static limit, where we follow \cite{Blau2018} Chap.23.2, put $C=0$, $D=1$ and take $ A, B $ profiles that depend only on $ r $. Such metric is diagonal: $g_{\mu\nu}=\diag(-A,B,r^2,r^2\sin^2\theta)$. The resulting energy-momentum tensor is also diagonal, its mixed component recording $T_\mu^\nu=\diag(-\rho,p_r,p_t,p_t)$ is expressed in terms of mass density $\rho$, radial pressure $p_r$ and transverse pressure $p_t$. Generally, these values are constrained by a model dependent equation of state (EOS), which can be written, e.g., as a dependence $p_{r,t}(\rho)$.

\begin{figure}
\begin{center}
\includegraphics[width=\textwidth]{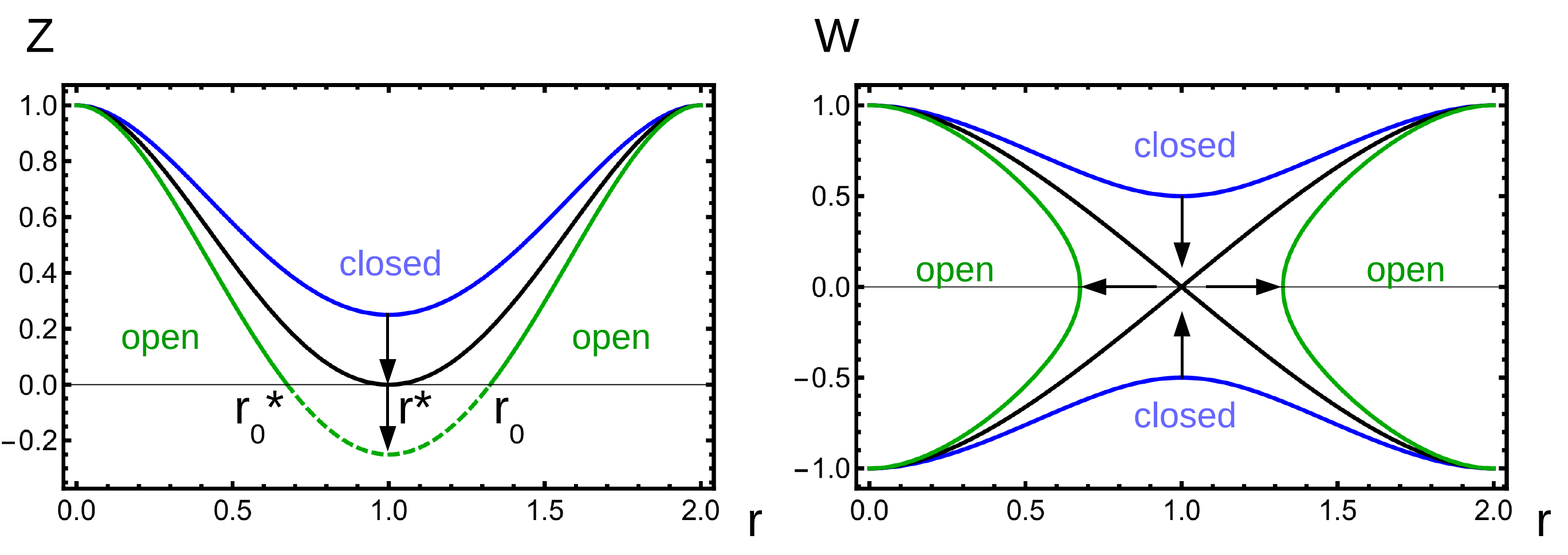}

\end{center}
\caption{The process of opening a wormhole. On the left -- in the variables $ Z (r) $, on the right -- in the variables $ W (r) = \pm \sqrt {Z (r)} $.}\label{f6}
\end{figure}

Conditions $ A> 0 $, $ B> 0 $ are imposed to select solutions without trapping horizons. The $A$-profile controls the time dilation and gravitational redshift effects, while the $B$-profile defines radial deformation of space and is related with the Misner-Sharp mass (MSM):
\begin{eqnarray}
&&M=r/2\ (1-B^{-1}).\label{msm0}
\end{eqnarray}
The throat of the wormhole corresponds to the minimal radius, where the following conditions are satisfied:
\begin{equation}
A(r_0)>0,\ B(r_0)\to+\infty,\ 2M(r_0)=r_0.\label{r0def}
\end{equation}
To obtain the wormhole (its symmetric variant \cite{Visser1996}), one connects the solution with its copy at the point $ r_0 $. Reparametrization can be used to obtain a globally regular solution: $ r \to L $, where the integral $ L = \pm \int_{r_0}^r \sqrt {B} dr $ represents the proper length and opposite signs are selected for different copies. The resulting function $ r (L) $ appears to be smooth and even, as well as $ A (r (L)) $ and other functions of it. 

After the application of the algorithm {\tt Einstein} to the metric (\ref{stdmetr}), we obtain
\begin{eqnarray}
&\rho=M'_r/(4 \pi r^2),\ p_r=(-2 A M + r (r - 2 M) A'_r)/(8 \pi r^3 A),\
p_t=(-r^2 (r - 2 M) (A'_r)^2 \label{rpMA1}\\
&+ 4 A^2 (M - r M'_r) + 2 r A (-A'_r (M + r (-1 + M'_r)) + r (r - 2 M) A''_{rr}))/(32 \pi r^3 A^2),\label{rpMA2}
\end{eqnarray}
this result can be also cross-checked against the formulas in \cite{Visser1996} and \cite{1604.02082}. Now let us introduce new variables
\begin{eqnarray}
&&Z=B^{-1},\ W=\pm\sqrt{Z},
\end{eqnarray}
here $B\to+\infty$ corresponds to $Z\to+0$, while the opposite signs at $ W $ correspond to different copies as considered above.

In $(Z,W)$-variables the suggested process of a wormhole opening is shown on Fig.\ref{f6}. The minimum of the function $ Z (r) $ goes down, passes through zero, and then goes into the negative region. Only the region $ Z \geq0 $ is physically used, and the negative part of the curve is shown by the dashed line in this figure. Taking the square root of this dependence, in the $ W (r) $ representation we see a typical hyperbolic catastrophe in which the upper and lower curves representing different sheets of space are reconnected. As a result, the wormhole is formed on the right side of the graph, a tunnel connecting different sheets of space. On the left, a bubble is formed, isolated from the outer space.

The formulas (\ref{rpMA1})-(\ref{rpMA2}) can be re-expressed in terms of the functions $ (A, Z) $ and their derivatives
\begin{eqnarray}
&\rho=-(-1 + Z + r Z'_r)/(8 \pi r^2),\ 
p_r=(A (-1 + Z) + r Z A'_r)/(8 \pi r^2 A), \label{rpAZ1}\\
&p_t=(-r Z (A'_r)^2 + 2 A^2 Z'_r + A (r A'_r Z'_r + 2 Z (A'_r + r A''_{rr})))/(32 \pi r A^2).\label{rpAZ2}
\end{eqnarray}
Analyzing the structure of this formulas, we see that these definitions are regular, i.e., $ (\rho, p_r, p_t) $ are finite if $(A,r)$ in the denominator are separated from zero ($>Const>0$) and if $ (A, Z) $ and their derivatives in the numerator are finite. For the deformation considered above, the $ Z $-profile is finite, with derivatives, as well as the $ A $-profile, which can be fixed and not changed. Thus, the topological reconnection described above can be performed in the class of finite matter terms.

In the next section and in Appendix~A we will continue this construction, with all necessary details. We will also take into account the dynamic terms and show that if the wormhole opening process is performed sufficiently slow (quasistatically), then the resulting matter terms turn out to be finite and described by the static expressions given here. A special consideration for the bifurcation point shows that the corresponding matter terms possess a mild type of singularity, equivalent to zero in distributional sense.

At the end of this section, we provide some more important formulas.

There are special conditions for opening a wormhole \cite{Visser1996,1604.02082} that result from the equations (\ref{rpAZ1})-(\ref{rpAZ2}), called {\it flare-out conditions}. In our model a special monotonous case $A'_r>0$ will be considered. Then, the throat of the wormhole satisfies the following conditions: 
\begin{eqnarray}
&r=r_0,\ Z=0,\ Z'_r>0\ \Leftrightarrow\ p_r=-1/(8\pi r^2)<0,\ \rho+p_r<0\label{flareout}\\
&\Rightarrow\ p_t>0,\ \rho+p_r+2p_t>0.\label{flareout2}
\end{eqnarray}
The conditions (\ref{flareout}) are necessary and sufficient for opening the wormhole, while the conditions (\ref{flareout2}) are their consequences (additional necessary conditions). In their derivation, the definitions (\ref{r0def}),(\ref{rpAZ1})-(\ref{rpAZ2}) were used, as well as the conditions $ A'_r(r_0)>0 $ and $ Z'_r (r_0)>0 $, according to Fig.\ref{f6}.

There are two additional special cases, one satisfied at the bifurcation point:
\begin{eqnarray}
&r=r^*,\ Z=0,\ Z'_r=0\ \Leftrightarrow\ p_r=-1/(8\pi r^2)<0,\ \rho+p_r=0\label{bif}\\
&\Rightarrow\ p_t=0,\ \rho+p_r+2p_t=0,\label{bif2}
\end{eqnarray}
the other corresponds to the maximal radius of the bubble:
\begin{eqnarray}
&r=r_0^*,\ Z=0,\ Z'_r<0\ \Leftrightarrow\ p_r=-1/(8\pi r^2)<0,\ \rho+p_r>0\label{bub}\\
&\Rightarrow\ p_t<0,\ \rho+p_r+2p_t<0.\label{bub2}
\end{eqnarray}

If the $A$-profile monotonicity condition is not strict, $ A'_r \geq0 $, the same conditions (\ref{flareout})-(\ref{bub2}) are met, except for one, $ \rho + p_r + 2p_t = 0 $ at the points where $ A'_r = 0 $.

\paragraph*{Comparison with other solutions.}
From the above formulas it can be directly seen that the quasistatic opening of the wormhole with nonzero $ r=r^* $ according to the scheme described here corresponds to the finite value of $ p_r $. An alternative scenario is often proposed, to take a wormhole of vanishingly small radius $r_0\to0$ and expand it to a sufficiently large size, see \cite {Visser1996} Chap.6.6.4, Chap.13.3, also \cite {Morris-etal}. This scenario requires unlimited pressure values. 

Fig.\ref{f7} shows different proposals for opening wormholes, where our solution of topology $ 2R ^ 3 \to R \times S ^ 2 + S ^ 3 $ is marked as (S1). Another option is considered in \cite {Visser1996} Chap. 9.1.1, \cite{GuthBubble}, also \cite {1407.6026} and references therein. This scenario (S2) also includes a bubble of space that seems to be inflated through a wormhole. Subsequently, the wormhole can be torn ($ r_0 \to0 $), and the bubble is completely separated from the outer space. Clearly, it is the other topological scheme $R^3\to R^3+S^3$, different from the one considered here. To break the wormhole in the quasistatic mode, this scheme will also require unlimited pressure values. There is also a related scenario (S3), see \cite{EuclideanWormholes}. It considers a process of separation and subsequent absorption of a baby universe. The corresponding embedding diagram looks like a handle, attached to a flat space, which is sometimes treated as a wormhole. Alternatively, reading the time slices on Fig.\ref{f7} (S3), one can see the process $R^3\to R^3+S^3\to R^3$, a combination of (S2) with its inverse, again, topologically different from (S1).

\paragraph*{Wormholes supported by Quantum Gravity.} It is widely known, that the creation of a wormhole requires an exotic type of matter. This follows directly from flare-out conditions, $ \rho + p_r <0 $ in the throat violates so called zero energy condition \cite{Blau2018} Chap.21.1. Note that here not the mass density is negative, as often suggested. The mass density can be positive, compensated by a large negative radial pressure. In general, the usage of energy conditions as selection rule for the solutions has been repeatedly criticized, see \cite{0205066} and references therein. It is also known that violation of energy conditions can occur in the framework of quantum gravity (QG).

The works \cite{0602086,0604013,0607039} have considered QG-corrections to Friedmann cosmological model with a scalar field. 

These corrections modify the mass density:
\begin{eqnarray}
&&\rho=\rho_{nom}(1-\rho_{nom}/\rho_{crit}),\label{rhoqg}
\end{eqnarray}
here $ \rho_{nom} $ is the classical nominal density, $ \rho_{crit} $ is the critical density of the order of Planck value, $ \rho_{crit} \sim \rho_P $. Thus, at the point where $ \rho_{nom}> \rho_{crit} $, the corrected mass density becomes negative, $ \rho <0 $. 

In \cite{1401.6562} and \cite{1409.1501}, the model of {\it Planck stars} has been constructed, based on this result. In this model, the stars exceed the critical density during their gravitational collapse, leading to gravitational repulsion and the {\it quantum bounce} effect: the black hole turns white, the collapse is reverted, the collapsing matter is ejected.

It is interesting to investigate the possibility that the effectively exotic matter terms created by quantum gravity do not lead to a quantum bounce, but to the formation of a wormhole. To study this topic below, we define the nominal density $ \rho_ {nom} $ and EOS of the form
\begin{eqnarray}
&&\rho=\rho(\rho_{nom}),\ p_r=p_r (\rho_{nom}),\ p_t=p_t (\rho_{nom}).\label{eosgen}
\end{eqnarray}
As we will see, the formation of a wormhole corresponds to EOS, in which, with increasing $ \rho_ {nom} $, the matter terms change their sign in a certain order. At first, the density becomes negative, after that -- the radial pressure, finally -- the transverse pressure. For this sequence, at a certain point, the flare-out conditions (\ref {flareout}), (\ref {flareout2}) are satisfied, leading to the wormhole opening.

\section{Opening wormholes}\label{sec3}

The possibility that a wormhole opens as a result of the effects of quantum gravity means that the solutions we study, in principle, can be formed naturally during the collapse of massive astrophysical objects. In addition, due to the coincidence of gravitational field in the outer range, the objects that we now consider as black holes can actually be wormholes. Below, we calculate a scenario in which a supermassive black hole in the center of the Milky Way galaxy is a wormhole. Thus, it will be shown that the model can be scaled to the parameters of real astrophysical objects. 

A similar scenario with the wormhole in the center of Milky Way was considered in \cite{0610441}. The difference is that in \cite{0610441} the wormhole was supported by exotic matter and magnetic field, while in our model it is supported by quantum gravity, and that in \cite{0610441} the wormhole was static, and in our scenario it is dynamic, can open and close.

\begin{figure}
\begin{center}
\includegraphics[width=\textwidth]{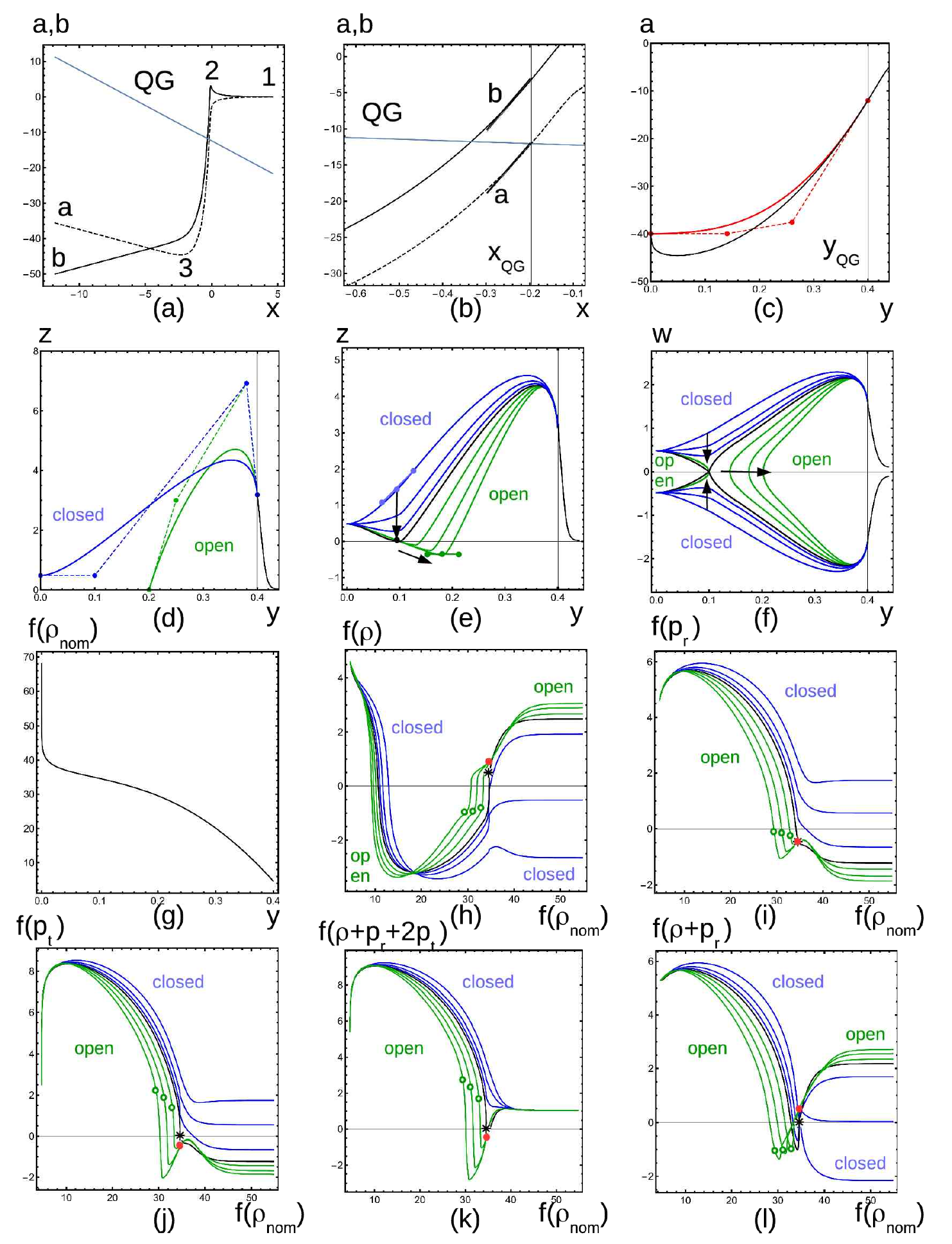}

\end{center}
\caption{Construction of dynamical QG wormhole (see text).}\label{f11}
\end{figure}

Now we are ready to calculate the matter terms necessary for the dynamic opening of a wormhole according to the scheme described above. The behavior of the main functions that characterize the solution is shown in Fig.\ref{f11} and in Tables~\ref{tab1},\ref{tab2}.

\paragraph*{Environment of the wormhole.} On its outer radius, the wormhole solution must be connected to the environment model, using boundary conditions for metric coefficients. To be able to interpret the solution in the context of quantum gravity, it is required that the nominal mass density in the environment model can reach Planck values.

As a suitable environment, we use the model of null radial dark matter (NRDM, \cite{1701.01569}). This model considers a static spherically symmetric distribution of dark matter, described by a perfect fluid EOS of the form $ \rho = p_r $, $ p_t = 0 $. The solutions of this model possess very high density and radial pressure in the central point:
\begin{eqnarray}
&&\rho=p_r=\epsilon/(8\pi r^2A).\label{rhopeff}
\end{eqnarray}
Here $ \epsilon> 0 $ is the scaling constant. In the central region of this solution, the redshift $A$-profile rapidly drops with a decreasing radius. In \cite{1701.01569} such behavior was called {\it red supershift}. It is closely related to the {\it mass inflation} phenomenon found in a black hole model with counterstreaming matter flows \cite{0411062}. This drop, together with the $ r^2 $ factor in the denominator, provides very high nominal mass density at the center, which grows rapidly and finally exceeds the Planck value.

\paragraph*{Coordinate transformations} are applied to display conveniently different scales of $ (A, B, Z, W, r) $ functions. While \cite{1701.01569} used logarithmic transformations:
\begin{eqnarray}
&&a=\log A,\ b=\log B,\ x=\log r,\label{abx}
\end{eqnarray}
we introduce the following scaling functions:
\begin{eqnarray}
&f(h)=\arcsinh(h/2),\ f^{-1}(h)=2\sinh(h),\ z=f(Z),\ w=f(W),\ y=f(r),\ l=f(L),
\end{eqnarray}
to display the vicinity of $ Z = 0 $, $ r = 0 $ points, connection of $ \pm W $ branches, and behaving logarithmically $ f (v) \sim \log v $ at $v\to\infty$.

The relations (\ref{rpAZ1})-(\ref{rpAZ2}) can be reformulated in terms of the above introduced functions using chain differentiation. As a result, material profiles $ (\rho, p_r, p_t) $ can be computed for the metric profiles $ (a (y), z (y)) $.

\paragraph*{QG cutoff} is defined as the point where the nominal density (\ref{rhopeff}) reaches the Planck density with an attenuation factor:
\begin{eqnarray}
&&\rho_{nom}=\epsilon/(8\pi r^2A)= \rho_P/N.\label{QGcut}
\end{eqnarray}
Being varied in the range $N \sim1-10$, this factor admits that QG effects start earlier than the exact Planck density value. Below we will consider a model example, where this factor is used to start QG effects at moderate density values.

After the QG cutoff point, NRDM EOS is changed to QG EOS of generic form (\ref{eosgen}). We proceed further setting the profiles $ (a (y), z (y)) $ as necessary to open the wormhole, then use (\ref{rpAZ1})-(\ref{rpAZ2}) to obtain the material profiles $ (\rho, p_r, p_t) $, defining a certain form of EOS. At first, we consider a model example with arbitrary chosen dimensions. Then, we perform the computation for the Milky Way's central black hole, showing that the model is scalable to real physical sizes.

\begin{table}
\begin{center}{\footnotesize
\caption{QG wormhole, model example}\label{tab1}

~

\def\arraystretch{1.1}
\begin{tabular}{l}
\br
Model parameters: $\epsilon=0.01$, $r_s=1$
\\ \mr
QG cutoff: $\rho_P/N=100$, $r_{QG}=0.821337$, $x_{QG}=-0.196821$, \\
$y_{QG}=0.399923$, $a_{QG}=-12.0409$, $b_{QG}=-3.1829$, $z_{QG}=3.18461$, \\ 
keypoints: $\{y_i,a_i\}=\{\{0.399923, -12.0409\}, \{0.259923, -37.5963\}, \{0.14, -40\}, \{0, -40\}\}$
\\ \mr
Closed state: $\{y_i,z_i\}=\{\{0.399923, 3.18461\}, \{0.13, 1.95\}, \{0.1, 1.5\}, \{0.379923, 6.92349\}, $ \\
$\{0.07, 1.05\}, \{0.03, 0.481212\}, \{0, 0.481212\}\}$, \\
redshift factor in the center $a(0)=-40$
\\ \mr
Open state: $\{y_i,z_i\}=\{\{0.399923, 3.18461\}, \{0.379923, 6.92349\}, \{0.212109, -0.38\}, $ \\
$\{0.182109, -0.38\}, \{0.152109, -0.38\}, \{0.03, 0.481212\}, \{0, 0.481212\}\}$, \\
redshift factor in the throat: $a(r_0)=-35.4799$,\\
radius of the throat: $r_0=0.406882$,\\
radius of the bubble: $r^*=0.200334$
\\ \br
\end{tabular}

}\end{center}
\end{table}

\begin{table}
\begin{center}{\footnotesize
\caption{QG wormhole in the center of Milky Way galaxy}\label{tab2}

~

\def\arraystretch{1.1}
\begin{tabular}{l}
\br
Model parameters: $\epsilon=4\cdot10^{-7}$, $r_{s,nom}=1.32\cdot10^{10}$m,
$r_2=r_s=1.1990455291886923\cdot10^{10}$m
\\ \mr 
QG cutoff: $N=10$, $\rho_P/N=3.82807\cdot10^{68}$m$^{-2}$, $r_{QG}=r_s-1.0003513617763519\cdot10^6$m,\\ 
$y_{QG}=23.207293345446693$, $a_{QG}=-222.2887073354903$, $z_{QG}=192.8253039716802$, \\ 
keypoints: $\{y_i,a_i\}=\{\{23.207293345446693, -222.2887073354903\},\{23.207283345446694,$\\
$-247.28371742005763\}, \{1, -250\}, \{0, -250\}\}$
\\ \mr
Closed state: $\{y_i,z_i\}=\{\{23.207293345446693, 192.8253039716802\},\{23.207193345446694, $\\
$442.77560481735327\},\{15, 300\}, \{10, 200\}, \{5, 100\},\{3, 0.48121182505960347\},$\\ 
$\{0, 0.48121182505960347\}\}$, \\
redshift factor in the center $a(0)=-250$
\\ \mr
Open state: $\{y_i,z_i\}=\{\{23.207293345446693, 192.8253039716802\},\{23.207193345446694, $ \\
$442.77560481735327\}, \{21.237780648029343, -0.2\},\{16.237780648029343, -0.2\}, $\\ 
$\{11.237780648029343, -0.2\},\{3, 0.48121182505960347\}, \{0, 0.48121182505960347\}\}$,\\ 
redshift factor in the throat: $a(r_0)=-241.7284225293921$,\\
radius of the throat: $r_0=1.3543177581795398\cdot10^7$m,\\
radius of the bubble: $r^*=22026.465749406787$m
\\ \br
\end{tabular}

}\end{center}
\end{table}

\paragraph*{QG wormhole, model example.} The solution with $ \epsilon = 0.01 $, $ r_s = 1 $ is shown on Fig.\ref{f11}a. The solution is found numerically, by integration procedure \cite{1701.01569}. 

The integration is performed with Mathematica {\tt NDSolve} algorithm, combining an automatic switching between fast explicit (e.g., {\tt ExplicitEuler}, {\tt ExplicitRungeKutta}, {\tt ExplicitMidpoint}) and stable implicit integration methods (e.g., {\tt Adams}, {\tt BDF}, {\tt ImplicitRungeKutta}, {\tt SymplecticPar\-titionedRungeKutta}), together with an adaptive {\tt DoubleStep} algorithm for the choice of integration step. In this algorithm, the error of integration is evaluated by Richardson's formula: $e=|y_2-y_1|/(2^p-1)$, where $y_{1,2}$ are the results of integration with a single $h$ step and two $h/2$ steps, $p$ is the order of the integration method. The step is adaptively selected to keep the estimated error in the range of specified precision tolerance, {\tt AccuracyGoal} and {\tt PrecisionGoal} in absolute and relative units. 

The parameter $ r_s $ denotes the nominal gravitational radius, used in the definition of a starting point. At point 1 the integration starts, the radius is chosen $ r_1 = 100 $, where the clock of the remote observer is set $ a_1 = 0 $ and $ b_1 $ for the given gravitational radius $ r_s $ is selected. Later, at point 2, the solution follows the Schwarzschild mode, with symmetrically raising $ b_2 $ and falling $ a_2 $. Since the profiles are given in a logarithmic scale, the metric coefficients $ A_2, B_2 $ differ by an order of magnitude from the initial values. Further, the red supershift starts, both metric coefficients rapidly fall, acquiring tens of orders of magnitude variation. Below this point several other structures appear, not important for our consideration, since they all are removed by QG cutoff. 

The QG cutoff is defined by the relation (\ref{QGcut}) and is shown by the straight line in Fig.\ref{f11}a. In the considered model example, a large attenuation factor is used, $ \rho_P / N = 100 $ in geometric units. Fig.\ref{f11}b shows a closeup of the QG cutoff, with the tangents related to the $ C^1 $-continuity of the profiles. Bézier curves are used to model the variation of the profiles, as shown on Fig.\ref{f11}c-f. Further details are explained in Appendix~A.

\paragraph*{The resulting EOS} in form of the $ (\rho, p_r, p_t) $ profiles is shown on Fig.\ref{f11}h-l. Each plot displays three profiles for the open states, three for the closed states and one separation line for the bifurcation. Open circles show the position of the wormhole throat $ y_0 $, filled circles show the maximal radius of the bubble $ y_0^* $, and the stars (sometimes coincident with filled circles) show the position of the bifurcation point $ y^* $. From the arrangement of these critical points, the validity of flare-out conditions (\ref{flareout}),(\ref{flareout2}), bifurcation conditions (\ref{bif}),(\ref{bif2}) and the bubble conditions (\ref{bub}),(\ref{bub2}) is directly visible. 

On the closed state lines of Fig.\ref{f11}h-l we see that flare-out conditions are not satisfied, in particular, $p_r$ becomes negative where $\rho+p_r$ is positive, so that the wormhole remains closed. Further, on open state lines, we see that in the range where $ \rho $ is negative, $ p_r $ also becomes negative. At the same time, $ p_t $ remains positive in this range. This leads to the appearance of zones where the flare-out conditions (\ref{flareout}),(\ref{flareout2}) are fulfilled, necessary and sufficient to open the wormhole. The wormhole is opened at the position $r$ where the equality $p_r=-1/(8\pi r^2)$ is satisfied. The right side of the graphs corresponds to the effect of separating the bubble. 

So far, the quasistatic opening of a wormhole in an NRDM environment has been investigated on a model example. The classical solution has been modified in the region of Planck densities. After computation of the EOS, we see that the material profiles become negative at large nominal density, similarly to the model of Planck stars. In our model, the material profiles change signs in a particular order. At first, the density, then the radial pressure, and finally the transverse pressure become negative. This opens a window of opportunities for the fulfillment of the flare-out conditions necessary and sufficient for opening the wormhole.

\paragraph*{QG wormhole in the center of Milky Way.} We repeat the computation, selecting the physical parameters of the Milky Way's (MW) central black hole. Table~\ref{tab2} presents the result of the calculation. Many significant digits are given in the table for the reproducibility of the calculation, which becomes very sensitive to the precision of Bézier parameters. All the plots have a similar structure to the ones in the considered model example.

The NRDM model is used in \cite{1701.01569} to represent the galactic dark matter distribution and the related rotation curves. In the single center approximation, where the dark matter distribution is related with the central black hole, the model provides flat rotation curves with the scaling parameter $ \epsilon = (v / c)^2 $. For MW rotation velocities $ v \sim200$~km/s it is $ \epsilon = 4 \cdot10^{-7 } $. The parameter $ r_s $ denotes the gravitational radius of the central black hole. Since the NRDM solution differs from the Schwarzschild one, the setting of the starting point at a large radius in weak fields should use a slightly different nominal value $ r_{s, nom} = 1.32 \cdot10^{10} $m, to be compatible with the observed value $ r_s = 1.2 \cdot10^{10} $m by Ghez et al. \cite{0808.2870}, in strong fields. We define the observed gravitational radius in the NRDM model as the point where the maximum of $ b $-profile is reached, $ r_2 = r_s $. All the local features of the gravitational field, such as the position of the photon sphere and the innermost stable circular orbits (ISCO), are defined by the strong field parameter $ r_s $, rather than by the weak field $ r_{s, nom}$.

As in the considered model example, after crossing the gravitational radius, the $ a $- and $ b $-profiles fall rapidly, while the density $ \rho $ rapidly increases, reproducing the red supershift and mass inflation effects. Passing 1000~km after $ r_s $, a very small distance relative to $ r_s $, the density comes into the Planck range. Here we use the factor $ N = 10 $ and perform the QG cutoff at $ \rho=\rho_P / N $. After that, the $ a $-profile in the classical NRDM model continues to fall, up to the values $\sim(-10^6) $, however, we replace it with a different profile that has the central value $ a = -250 $. Also, a different profile is selected for $ z $, making opening and closing of the wormhole possible. For the selected parameters, the bifurcation point is located at $ r^* = 22$~km. Here the wormhole throat and the bubble of this radius are formed. Further, in the considered scenario, the bubble radius remains constant and the wormhole throat radius increases, finally reaching $ r_0 = 13543$~km.

\section{Stargates and teleporters}\label{sec4}

\begin{figure}
\begin{center}
\includegraphics[width=0.6\textwidth]{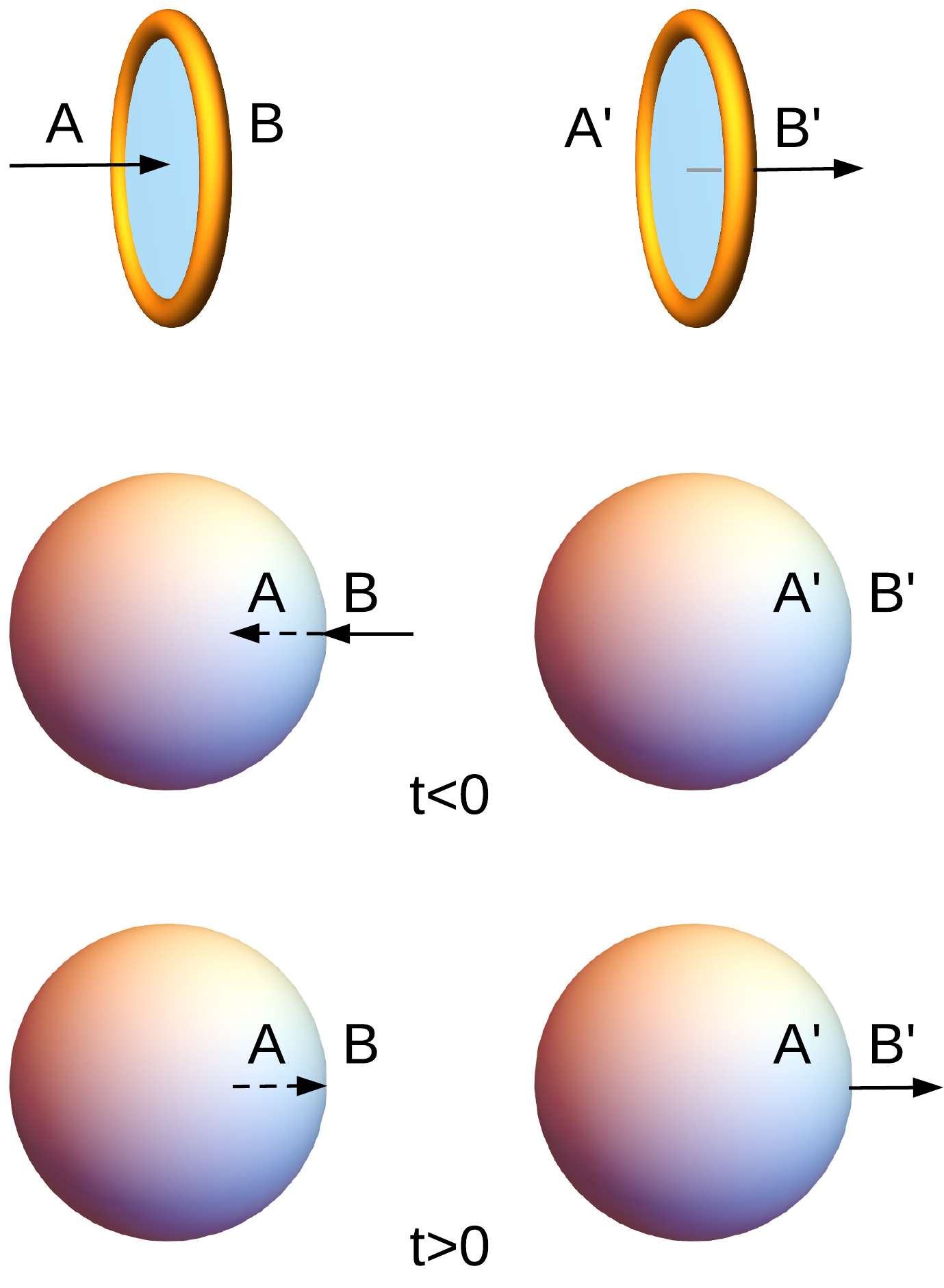}

\end{center}
\caption{On the top -- stargate solution, in the center and on the bottom -- teleporter solution.}\label{f8}
\end{figure}

\vspace{2mm}
\paragraph*{Stargate} solution is a special type of wormhole, described in several variants in the book by Visser \cite {Visser1996}, on the basis of his original works \cite {Visser1, Visser2}. A schematic diagram of such solutions is shown in Fig.\ref{f8} at the top. Consider two copies of space $ R ^ 3 $, where in each copy the disk $ D_ {1,2} $ is cut out. Further, the space adjacent to the different sides of the disks is glued crosslike, according to the scheme $ AB '$, $ BA' $. In practice, when using a stargate, the traveler moves from the $ A $ side of the disk $ D_1 $ in one universe to the $ B '$ side of the disk $ D_2 $ in another universe, or in a remote region of the same universe.

\paragraph*{Teleporter} is a new type of solution we will construct in this section. It is depicted in Fig.\ref {f8} in the center and at the bottom. Two spheres $ S ^ 2 $ are cut out in two copies of the space $ R ^ 3 $. In the time interval $ t <0 $, the space adjacent to the spheres inside and outside is glued in the direct way $ AB $, $ A'B '$, and in the interval $ t> 0 $ in the crosslike way $ AB' $, $ BA '$. At $ t <0 $, the traveler crosses the sphere in one universe via the $ BA $ connection, then, at $ t> 0 $, crosses the sphere in the $ AB '$ connection, in another universe, or in a remote part of the same universe.

First of all, we note that the concept of teleportation as instantaneous movement over long distances easily fits into the general theory of relativity. The key is the presence of topologically non-trivial solutions. Wormhole is a shortcut between points that can be located many light years apart. The teleporter is also such a shortcut, being just a wormhole of a different type, obtained using another procedure of cuts and identifications.

\paragraph*{Duality transformations.} In more detail, we show that teleporter geometry is related with other known solutions by certain duality transformations. To start with, it is topologically dual to the opening wormhole scheme discussed above, in which, for $ t> 0 $, the second possible gluing option $ AA '$, $ BB' $ is used. For this solution, at $ t> 0 $, the traveler crosses the wormhole throat along the connection $ BB '$ towards the destination, while the connection $ AA' $ describes the separated bubble.

\begin{figure}
\begin{center}
\includegraphics[width=0.8\textwidth]{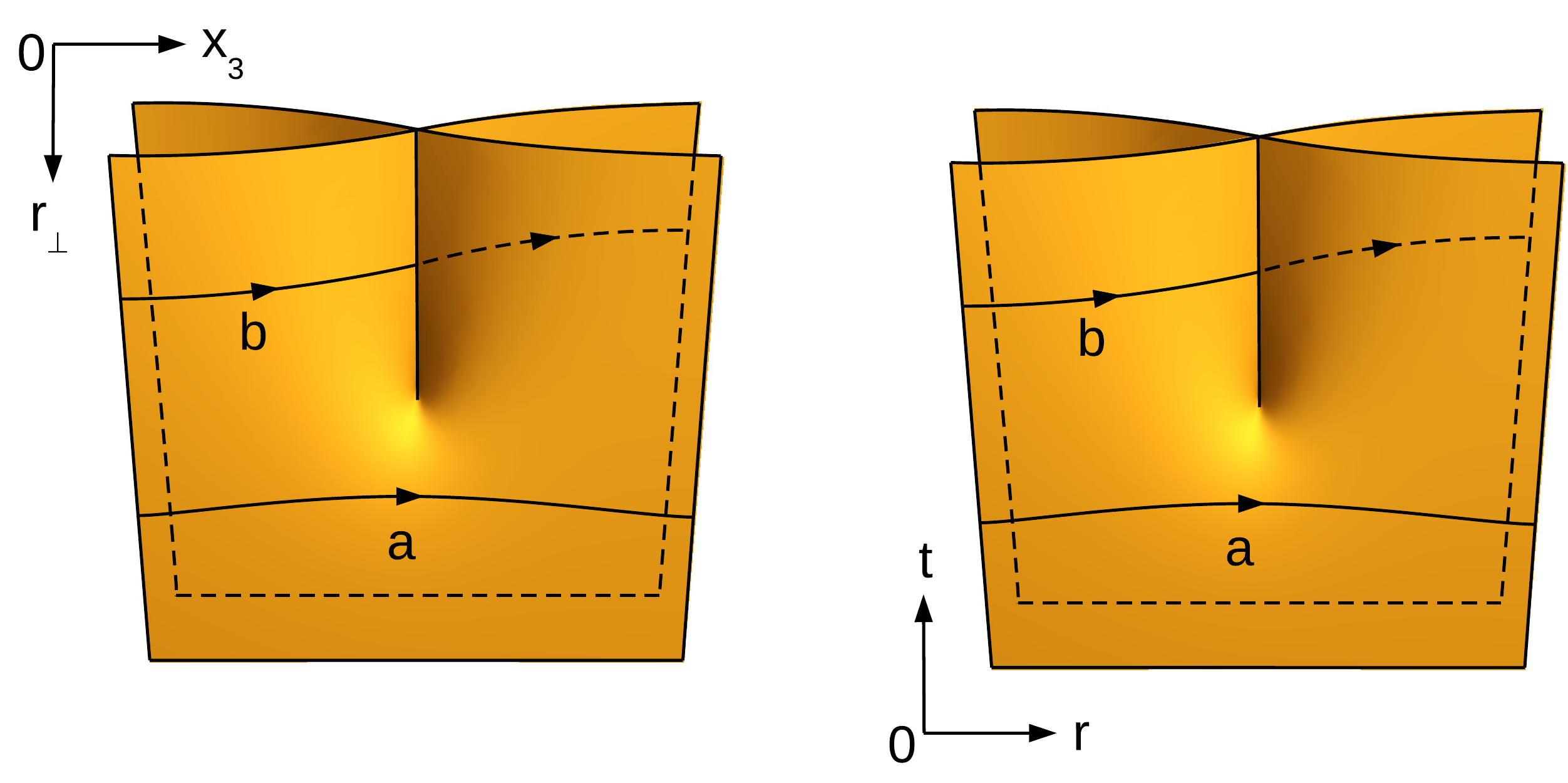}

\end{center}
\caption{Embedding diagrams, on the left -- stargate, on the right -- teleporter.}\label{f9}
\end{figure}

Further, Fig.\ref {f9} shows the embedding diagrams for stargate and teleporter. It is noteworthy that both diagrams are represented by the same surface with two sheets and a branching point. They are just represented in different coordinate systems. The stargate in the figure on the left is in $ (r_ \perp, x_3) $, composed of the components of a cylindrical coordinate system of 3-dimensional space. The teleporter in the figure on the right is in $ (r, t) $, where $ r $ is the radius of the spherical coordinate system, $ t $ is the time. Below, in Appendix~B, we show that the geometrical duality of stargate and teleporter solutions can be extended to the algebraic duality. Namely, these solutions are related by a special transformation, known as Wick rotation, a formal replacement of time by a complex value $t\to ix_3$.

Fig.\ref {f9} also shows the paths of the traveler described above. In the figure on the left, path (a) corresponds to $ r_ \perp> a $, where $ a $ is the radius of the disk, along this path the traveler does not cross the disk and remains in one universe. Path (b) corresponds to $ r_ \perp <a $, the traveler crosses the disk and moves to another universe. In the figure on the right, the path (a) corresponds to $ t <0 $ and remains in one universe, the path (b) corresponds to $ t> 0 $ and goes to another universe.

\paragraph*{Distribution of matter.} The third coordinate in Fig.\ref{f9} is needed only for visualization. If this coordinate is flattened to zero, the surfaces turn out to be flat, the metric induced on them will be also flat ($g_{\mu\nu}=\diag(-1,1,1,1)$) almost everywhere. An exception will be the branching point at which the metric has a defect. This means that for both geometries under consideration the matter term will be equal to zero (empty space) everywhere except for the branching point. 

For a stargate, this means that the matter will be concentrated on the perimeter of the disk, at all time instants. The stargate geometry is stationary, and the matter term will be localized on the cylinder $ S ^ 1 \times R ^ 1 $ in spacetime. For the teleporter, the matter will be concentrated in the vicinity of the sphere $ S ^ 2 $ at the moment $ t = 0 $, that is, on a manifold of the same dimension, but of a different topology. Teleporter geometry is essentially dynamic, and is flat almost everywhere, except for a sphere arising for one instant of time. Both geometries are traversable in the sense of \cite {Visser1996}, the traveler can move from one universe to another without crossing regions of high curvature and concentration of (exotic) matter. In the first case, the traveler should avoid intersections with the stargate perimeter, in the second -- with the teleportation sphere at the time of transition.

In Appendix~B, we calculated the matter terms for the geometries under consideration. For the stargate, there is an exotic string on the perimeter coiled into a ring, with negative linear density and negative tension equal to $ \mu = \tau = -1 / 4 $ in geometric units ($G=c=1$), that is, about minus 0.2 Jupiter mass per meter in natural units. The result is the same as in \cite {Visser1996,Sorkin2, 9305009}, where other methods were used for its derivation.

\begin{figure}
\begin{center}
\includegraphics[width=\textwidth]{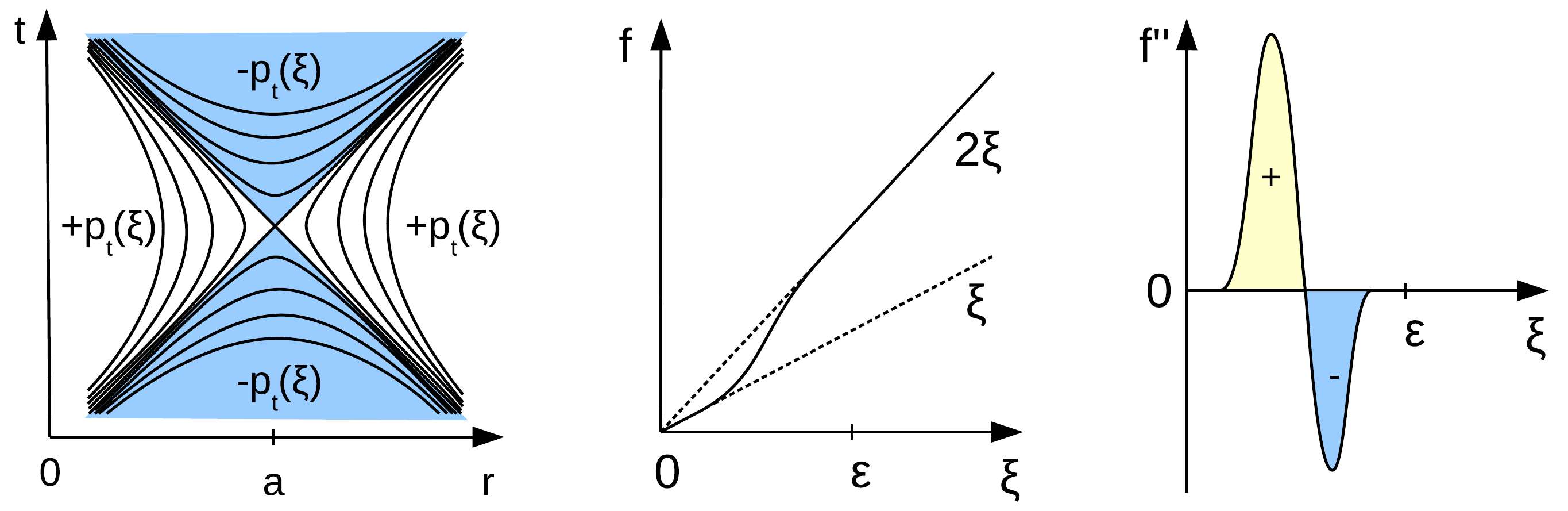}

\end{center}
\caption{Distribution of matter for teleporter. On the left -- in $ (r, t) $ coordinates, in the center -- regularization function $f(\xi)$, on the right -- the function $ p_t(-\det g)^{1/2}\sim f''(\xi) $.}\label{f10}
\end{figure}

The matter term for the teleporter has a different structure, see Fig.\ref{f10}. The active part of energy-momentum tensor is given by two components of the transverse pressure $p_t$. In the coordinates $ (r, t) $, the distribution of $p_t$ has alternating sign $\eta=\pm1$ in the sectors shown in the figure. Altogether there are 8 of such sectors, 4 in one and 4 in another universe. The distribution is constant on the hyperbolic orbits marked by the parameter $ \xi $, which is similar to the radius in the polar coordinate system. The computation uses a regularization function $f(\xi)$, that smoothly interpolates between $\xi$ and $2\xi$ dependencies in the interval $[0,\epsilon]$. The values multiplied to $(-\det g)^{1/2}$ in general relativity are called densitized values. The densitized pressure is proportional to the second derivative of the regularization function $p_t(-\det g)^{1/2}\sim f''(\xi)$, with the proportionality coefficient of the same order as for stargate geometry. 

Further, in Appendix~B, we prove that when the regularization is removed, $\epsilon\to0$, the function $f''(\xi)$ tends to $\delta(\xi)$. However, due the alternating sign, the function $f''(\xi)/\eta$ on the plane tends to zero in the distributional sense. In other words, after multiplication to a test function from an appropriate class and integration over the coordinate volume, it tends to zero when the regularization is taken off. 

\paragraph*{Comparison with other solutions.} Although the result is equivalent to zero as a generalized function, it is not the same as the usual function tending to zero. For example, this function cannot be squared, this will lead to an infinite result after removing the regularization. Let us compare this case with a well known problem of a toroidal compactification $T^n$, i.e., a flat space in a box, whose opposite sides are identified (see \cite{Visser1996} Chap.17.3.2). Although possessing some similarity with our identification procedure for the teleporter, in this case the matter term vanishes identically as a function. Therefore, because of all these subtleties, we prefer to describe the structure of the regularized solution, representing a physical approximation to an idealized geometry, obtained in the limit $\epsilon\to0$.

Comparing our result with other works on topology change and degenerate metrics in general relativity, the authors of \cite{Yodzis1972, 9711069, Horowitz1991, 9109030} agree that these cases generally correspond to a mild type of singularity. A special opinion is \cite{Sorkin2}, where a complex regularization for Morse singularity of saddle type is used, adding a small imaginary term $i\gamma$ to the metric (\ref{morsemetr}). Quite interesting is that the result for the densitized scalar curvature in this case is also complex, $4\pi i\delta_2(x,y)$, after removal of the regularization. Although this can have a profound meaning in quantum theory \cite{Sorkin2}, we consider the classical theory and prefer to stay with real-valued expressions for the metric and Einstein tensor.

\paragraph*{What are the stargate perimeter and teleportation sphere made of?} In any case, the matter is exotic, violates the energy conditions, as well as the matter necessary for constructing of static wormholes. For the stargate, this is an exotic string of negative mass and tension. It differs from commonly used positive mass and tension cosmic strings. As noted in \cite {Visser1996} Chap.15.3.1, their actions are proportional to each other, and therefore geometrically such solutions coincide. However, due to the difference in the sign of matter term, they produce different gravitational fields. In particular, strings with positive mass and tension produce a specific deformation of space, known as deficit of angle. The strings with negative mass and tension produce a negative deficit (an excess), in particular, a doubling of angle necessary for solutions with the branching point of stargate type. These deformations are considered in more detail in Appendix~B.

For the teleporter, the matter term of the regularized solution has only transverse pressure, but no radial pressure and no mass. Several possibilities for creating such matter distribution can be considered. (i) A gas consisting of two components: $ (\rho, p_r, p_t) $ and $ (- \rho, -p_r, 0) $, summing up to the matter with purely transverse pressure $ (0,0, p_t) $. (ii) Tachyons (hypothetic particles with spacelike worldlines). Every point in the diagram of Fig.\ref{f10} on the left is a sphere existing for one instant of time. On each sphere, draw a system of great circles, each is a closed worldline of the tachyon on the sphere. Thus, a two-dimensional tachyon gas is placed on the sphere, creating the necessary transverse pressure. The sign of this pressure is regulated by a mass factor common to all tachyons, which can be either positive or negative. (iii) A string coiled into a ring and existing for one instant, having zero mass and non-zero (positive or negative) tension. A sphere with transverse pressure can be assembled from such strings, and from such spheres -- an alternating sign distribution necessary for operation of the teleporter.

\section{Conclusion}\label{sec5}

In general relativity, two examples of solutions of variable topology are constructed. The first corresponds to the dynamic opening of a wormhole according to a scheme of a new type, in which the initial state contains two copies of three-dimensional space and the final state contains a closed bubble (baby universe) and two copies of three-dimensional space connected by a wormhole. The second example represents a topologically dual scheme of maps gluing, it corresponds to an instant swap of two volumes in space, which can be interpreted as a teleportation event. The second example is also related to the previously investigated stationary solution of a stargate (dihedral wormhole) type, which has the same embedding diagram, in different coordinates, and is related to the new solution by Wick rotation.

For both solutions, the corresponding matter distributions are calculated. For the first solution, the matter terms turn out to be finite, except for the immediate vicinity of the bifurcation point, where a mild singularity of Morse saddle type is located. For the second solution the matter terms are concentrated near the teleportation sphere, similar to a stargate, in which the matter terms are concentrated on the perimeter. For both solutions, the bifurcation point of wormhole opening and the branching point of teleporter represent a sign alternating singularity, equivalent to zero in distributional sense.

Although the matter composition in all considered solutions is exotic, violating the energy conditions, there is a principal possibility of creating such solutions via quantum effects. Similar processes were previously considered in the Planck stars model, in which quantum gravity corrections led to effectively negative mass density, repulsive force and the quantum bounce phenomenon. We calculated a scenario in which similar repulsive terms do not lead to a quantum bounce, but to the dynamic opening of a wormhole. Such scenario can be scaled to real astrophysical sizes corresponding to the central black hole in the Milky Way, describing the principal possibility of opening a wormhole as a result of natural astrophysical phenomena.

For solutions of stargate and teleporter types, several matter composition options were considered, in form of an exotic string coiled into a ring, a string with zero-mass and non-zero tension, a two-dimensional tachyon gas, and a two-component gas of a normal and exotic type. In particular, the combination of a normal matter with exotic matter from the core of Planck stars gives a principal possibility for engineering such solutions.

\section*{Acknowledgment}

The author thanks Kira Konich for proofreading the paper. 

\footnotesize

\appendix
\section{Opening wormholes, details of construction}

\vspace{3mm}\paragraph*{Setting a(y) profile.} 
The modeling of $ a (y) $ profile is shown in Fig.\ref{f11}c. Bézier curves are used for the definition of profiles. The Bézier curve is defined by a set of $ n + 1 $ control points $ \{P_0, P_1, ..., P_n \} $:
\begin{equation}
X(t)=\sum_{i=0}^n P_i\, B_{i,n}(t),\ B_{i,n}(t)=C^n_i\, t^i(1-t)^{n-i},
\end{equation}
(see, e.g., mathworld.wolfram.com/BezierCurve.html), where $ t \in [0,1] $, $ B_{i, n} (t) $ are Bernstein polynomials, $ C^n_i $ are binomial coefficients. $ P_0 $ is the starting point of the curve, $ P_n $ is the end point, $ P_1-P_0 $ and $ P_n-P_{n-1} $ define tangent vectors to the curve at these points. The curves of complex shape are modeled with the aid of several Bézier curves, $ C^1 $-smoothly patched together. For the definition of $ a (y) $ and $ z (y) $ profiles, we use Bézier curves with $ n = 3 $. Simple profiles Fig.\ref{f11}c,d are modeled with 4 control points $ \{P_0, P_1, P_2, P_3 \} $, more complex ones Fig.\ref{f11}e -- with 7 points, taking $ \{P_0, P_1, P_2, P_3 \} $ for the first patch of Bézier curve and $ \{P_3, P_4, P_5, P_6 \} $ for the second patch. 

The formulas (\ref{rpAZ1})-(\ref{rpAZ2}) use the second derivative of the $ a $-profile in $ p_t $, thus, a $ C^1 $-connection of the $ a $-profile creates a discontinuity in $ p_t $. This discontinuity is physically admitted, its meaning is a sharply beginning transverse interaction between the radial flows of dark matter, happening beyond the QG cutoff point $ y_{QG} $. If $ C^2 $-connection of Bézier curves is enabled, it will lead to a continuous rapid increase of $ p_t $, with the same physical meaning. Therefore, there is no physical difference between $ C^1 $ and $ C^2 $ connections for the $ a $-profile.

Next, by choosing the control points, we achieve that $ a (y) $ is a monotonically decreasing function with decreasing $ y $. From here and the formula (\ref{QGcut}), we see that $ \rho_{nom} (y) $ is monotonically growing with decreasing $ y $. Fig.\ref{f11}g shows this function using the scaling map $ f (\rho_{nom} (y)) $. The monotonicity property is convenient in our constructions, since the resulting dependencies $ (\rho, p_r, p_t) (y) $ are easy to transcode $ y \to \rho_{nom} $ and get the EOS in the desired form, as a function of $ \rho_{nom} $. We also note that $ a (y) $-profile can change during evolution, but for simplicity we consider a particular class of solutions with a fixed profile $ a (y) $.

\paragraph*{Setting z(y) profile.} Fig.\ref{f11}d shows the $ z (y) $ profile for two wormhole states, closed and open. At the QG cutoff point, the $ C^1 $-connection provides a common tangent vector with the classical region. As a result, the $ z (y) $-profile continues to increase with decreasing $ y $, corresponding to the decreasing MSM and positive mass density, similar to the NRDM model. Then, the $ z (y) $-profile starts to decrease. Here the QG effects come into play, making the mass density negative. Then, the open and closed curves behave differently.

For the closed state, the point $ y = r = 0 $ is reachable. Also, we consider solutions with a density bounded near the point $ r = 0 $. Thus, in the vicinity of the center we have:
\begin{equation}
M \sim4 / 3\, \pi \rho (0) r^3,\ Z = 1-2M / r \sim 1-8 / 3\, \pi \rho (0) r^2, 
\end{equation}
therefore, $ z (y) $ at $ y \to0 $ tends to $ z (0) = f (1/2) \approx0.481212 $ with a tangent directed horizontally. If the condition $ \rho (0) <0 $ is met, typical for the negative density exotic core, then the profile will approach this tangent from above, see Fig.\ref{f11}d.

For the open state, the profile goes to $ z \to + 0 $, as typical for the wormhole throat. The radius of the throat is controlled by the parameter $ y_0 $.

Now we construct a continuous transition from the closed curve to the open curve in Fig.\ref{f11}d, while maintaining the above boundary conditions. It consists of two phases, shown in Fig.\ref{f11}e.

At first, a bending of the $ z (y) $-profile happens, corresponding to an inner positive density core inside a larger negative density core. Then the minimum value $ z = 0 $ is reached at the point $ y^* $. Here, a bifurcation happens, leading to the wormhole opening.

Secondly, the $ z (y) $-profile goes deeper into the negative region. While physically only the part  $ z \geq0 $ is used, the control points and the part of Bézier curve go into the negative region. This negative part just fills the gap in the resulting EOS, which really is never accessed by the solution.

Fig.\ref{f11}e depicts a topological reconnection, where the wormhole suddenly appears at some nonzero radius, while the inner space below this radius becomes connected to its copy, forming a closed bubble. This bubble is totally disconnected from the outer space, thus, the observers confined in it will not have a way to return, other than performing a reverse procedure, connecting the bubble with the wormhole of the appropriate size.

The diagram in Fig.\ref{f11}e describes a topological rearrangement in which a wormhole forms suddenly at a nonzero radius, and the space inside this radius sticks together with its copy on the boundary sphere and forms a closed bubble that is completely disconnected from the external space. Matter and observers there will no longer be able to return to the outer space, except as a result of the reverse procedure of gluing the bubble to a suitable wormhole.

Below we consider a special type of the bubble evolution, when, after the formation, the bubble preserves its external radius. In quasistatic limit, it corresponds to the case when MSM of the bubble is conserved: 
\begin{equation}
Z (r^*_0) = 1-2M (r^*_0) / r^*_0 = 0\ \Rightarrow\ M (r^*_0) = r^*_0 / 2 = Const.
\end{equation}
Generally, it will not be the case, an arbitrary evolution of $Z,W$-profiles, not supporting this constraint, is also allowed. In particular, the bubble can evolve according to closed scenarios of Friedmann cosmological model. (Conservation of MSM and bubble radius depends on various model details, in particular: (i) $C=0$, $D=1$ in definition of metric, (ii) energy flow through the boundary of the system, (iii) the work of internal pressure forces. See the detailed discussion in \cite{Blau2018} Chap.23.6, Chap.34.4-7.) We use the freedom of our construction and fix a special stationary $r^*_0$ scenario for the bubble. This choice does not influence the dynamics of the wormhole.

Let us consider the bifurcation point in more detail. The wormhole throat possesses a positive mass $ M (r^*) = r^* / 2 $, see \cite{Visser1996}. There is no physical density creating this mass, it appears just due to the $ Z (r^*) = 0 $ boundary condition. However, at the moment preceding the bifurcation, such physical mass was located below a $ r^* $ radius. After the bifurcation, in the considered scenario, this mass becomes located in the bubble and is further preserved, while the radius of the wormhole throat starts to increase.

Fig.\ref{f11}f depicts the evolution of the $ w (y) $-profile. Here we clearly see the hyperbolic reconnection, corresponding to the opening of the wormhole and the formation of the bubble. While the bubble profile in further evolution just slightly changes its shape, the wormhole throat radius continues to increase towards its final $ r_0 $ value.

\paragraph*{Dynamical terms in vicinity of the throat.} So far, we have looked at a sequence of static configurations, interpreted as a quasistatic wormhole opening process. A special consideration is required in the vicinity of a throat $r_{min}$ and at a maximal radius of the bubble $r_{max}$, which will be analyzed now, as well as in the vicinity of the bifurcation point $r^*$, which will be analyzed later.

Specifying the behavior of $r$-coordinate in the vicinity of the throat:
\begin{equation}
r(t,L)=r_0(t)+c(t)L^2,
\end{equation}
where $r_0(t)$ defines the evolution of the throat radius, $c(t)>0$ corresponds to $r>r_{min}$ throat and $c(t)<0$ -- to $r<r_{max}$ bubble, we have
\begin{eqnarray}
&&L(r,t)=((r-r_0(t))/c(t))^{1/2},\ ds^2=-A dt^2 +dL^2+r^2d\Omega^2\\
&&=(-A+(L'_t)^2)dt^2+(L'_r)^2 dr^2+2 L'_t L'_r dt dr +r^2d\Omega^2.
\end{eqnarray}
Pay attention to the appearance of non-diagonal term in the metric. Note also that $L'_t$ and $L'_r$ contain a singular multiplier $|r-r_0|^{-1/2}$. The isolation of singular terms shows $g_{tr}\sim -r_0'/(4c(r-r_0))$ and $g_{tt}\sim(r_0')^2/(4c(r-r_0))$, the terms that remain active in a small vicinity $r\sim r_0$, even if $r_0'$ becomes arbitrarily small.

The application of {\tt Einstein} algorithm leads to a lengthy expression for $G_\mu^\nu$ with the following properties: (i) denominators of components are monomials of $(r,A,c)$; (ii) numerators are polynomials of $(r,A,c,r_0)$ and their derivatives up to the second order; (iii) all components are finite, provided that the denominator values are separated from zero and numerator values and derivatives are finite; (iv) when time derivatives vanish, the expressions coincide with those of quasistatic limit (\ref{rpAZ2}). In addition:
\begin{equation}
Z=(L'_r)^{-2}=4c\,(r-r_0),\ \det g=-(A/Z)\,r^4\sin^2\theta,
\end{equation}
i.e., $Z\to0$ at $r\to r_0$ linearly, while $(-\det g)^{1/2}\sim|r-r_0|^{-1/2}$ contains an integrable singularity. As a result, the densities $T_\mu^\nu(-\det g)^{1/2}$ are all integrable.

We have also tested that the omission of $L_t'$ terms in the metric produces Einstein tensor with strong singularities.

\paragraph*{The bifurcation point.} According to \cite{Sorkin2}, the coordinates in the vicinity of singular Morse point of saddle type can be selected so that Morse function will take a canonical form $t=x^2-y^2$. Then the metric (\ref{morsemetr}) will be
\begin{equation}
ds^2=(x^2+y^2-\zeta x^2)dx^2+(x^2+y^2+\zeta y^2)dy^2+2\zeta xy dx dy.
\end{equation}
Using the coordinate transformation \cite{Sorkin2}, the metric is simplified to
\begin{equation}
u=(x^2-y^2)\sqrt{\zeta-1}/2,\ v=xy,\ ds^2=-du^2+dv^2,
\end{equation}
coincident with the flat Minkowski plane, covered twice by the transformation $(x,y)\to(u,v)$. 

This type of singularity will be considered in details below. The result is that the matter term vanishes everywhere, except for the origin, where a mild singularity is located, equivalent to zero in distributional sense. 

\section{Stargates and teleporters, details of construction}

In this section, we calculate the distribution of matter for the spacetime geometry shown in Fig.\ref{f9}. The calculation for the two cases under consideration has a similar form, the difference is only the Euclidean vs Lorentzian signature of the metric and the use of a cylindrical vs spherical coordinate system. The difficulty is that in both cases spacetime is locally flat, and the standard algebraic calculation produces zero matter term. In fact, this calculation is valid everywhere except for the origin, and to obtain a matter term concentrated at the origin, a special regularization is required.

\begin{figure}
\begin{center}
\includegraphics[width=0.3\textwidth]{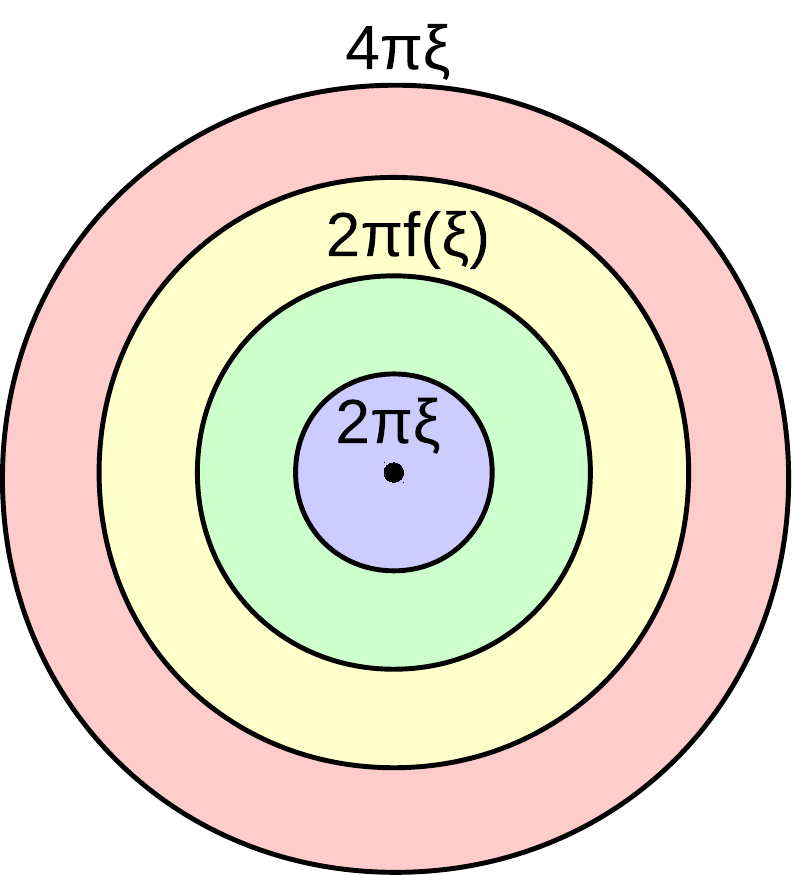}
~~~\includegraphics[width=0.6\textwidth]{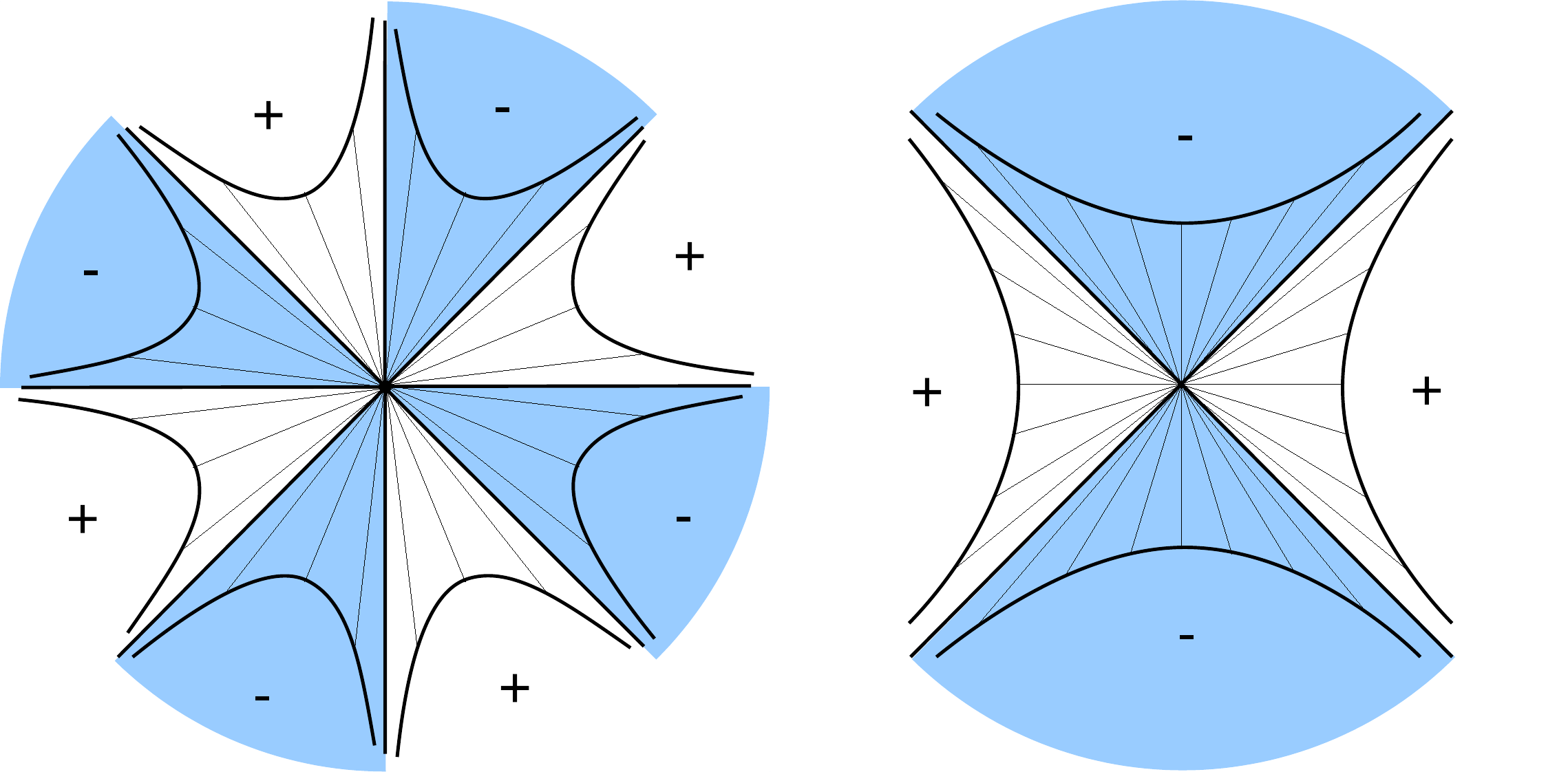}

\end{center}
\caption{Construction of a stargate (on the left) and teleporter (in the center and on the right), in a double cover polar coordinate system.}\label{f12}
\end{figure}

\paragraph*{Stargates.} The double cover geometry in Fig.\ref {f9} represents so called conic singularity with defect angle $(-2\pi)$, with the known answer for Einstein tensor \cite{Visser1996,Sorkin2,9907105}. Our purpose is to obtain this result with a physically based regularization, which can be also transferred to the Lorentzian case. The final result will not depend on the regularization function. See also \cite{9305009} for similar ideas.

The double cover geometry can be represented in the polar coordinate system $ (\xi, \alpha) $, shown in Fig.\ref{f12} on the left. The deviation of the metric from the flat Euclidean one is that the circumference here is $ 4 \pi \xi $, doubled compared to the flat case. The main idea of the proposed calculation method is to interpolate the circumference value in the form $ 2 \pi f (\xi) $, where the smooth function $ f (\xi) $ is $ 2 \xi $ at large distances, outside $ \xi <\epsilon $ neighborhood, and is equal to $ \xi $ near $ \xi = 0 $. Near the origin, the space will turn out to be Euclidean flat, and the concentrated matter term assigned to the origin will be equal to zero. Thus, the constructed regularization displaces the matter concentrated at the origin to the $ \epsilon $-neighborhood, where its distribution can be calculated by the standard method. The localization of matter is controlled by the parameter $ \epsilon $, which can go to zero at the end of the calculation.

We will increase the complexity of the problem gradually and first consider 3-dimensional case:
\begin{equation}
n=3,\ x=(t,\xi,\alpha),\ g=\diag(-1,1,f(\xi)^2), \label{sg3}
\end{equation}
where $ x $ is the chosen coordinate system, $ g $ is the metric tensor, $ f (\xi) $ is the regularizing function introduced above. The application of the algorithm {\tt Einstein} leads to the tensor with one nonzero component
\begin{equation}
G^t_t=-8\pi\rho=f''(\xi)/f(\xi), \label{sg3G}
\end{equation}
where $ \rho $ is the mass density. Note that after inserting $(-\det g)^{1/2}=f(\xi)$ to the integration measure, the total mass integral over the $ \epsilon $-neighborhood has the form
\begin{equation}
\int \rho f(\xi)d\xi d\alpha =(-1/4)\int_0^\epsilon f''(\xi)d\xi =(-1/4)(f'(\epsilon)-f'(0))=-1/4. 
\end{equation}
It is noteworthy that the result does not depend on the specific form of the regularizing function, only on the boundary values of its derivatives, which we fixed on $ f '(\epsilon) = 2 $, $ f' (0) = 1 $ by construction above. Thus, $f''$ plays the role of a regularized delta function. (Note: since $f''$ is not everywhere positive in our construction, this property is non-trivial and will be proven strictly below).

Now consider 4-dimensional case, an infinite string extended along z-axis:
\begin{equation}
n=4,\ x=(t,z,\xi,\alpha),\ g=\diag(-1,1,1,f(\xi)^2). 
\end{equation}
In this case, we obtain the Einstein tensor with two nonzero components
\begin{equation}
G^t_t=G^z_z=f''/f.
\end{equation}
Here, after multiplication by $(-\det g)^{1/2}=f$, the second derivative $f''$ again plays the role of a regularized delta function, and the components $ G ^ t_t = -8 \pi \rho $, $ G ^ z_z = 8 \pi p_z $ correspond to compensating one another negative mass density and positive longitudinal pressure, $ \rho + p_z = 0 $. After integration over the $ \epsilon $-neighborhood, we get the linear mass density of the string and its tension (that is, minus pressure force):
\begin{equation}
\mu=\tau=-1/4,
\end{equation}
in full accordance with the formula (15.24) \cite {Visser1996}.

Next, in the same dimension, we coil the string into a circle of radius $ a $, getting the stargate geometry:
\begin{equation}
n=4,\ x=(t,\xi,\alpha,\phi),\ g=\diag(-1,1,f(\xi)^2,r_\perp^2),
\end{equation}
here $ (\xi, \alpha) $ represent a double cover polar coordinate system on the plane $ (r_ \perp, z) $, which is a subset of the standard cylindrical coordinate system in 3-dimensional space $ (r_ \perp, z, \phi) $. Interpolation is used for
\begin{equation}
r_\perp=a+(2\xi-f(\xi))\cos\alpha+(f(\xi)-\xi)\cos 2\alpha,
\end{equation}
providing a smooth transition from single to double coverage of the plane $ (r_ \perp, z) $, from a locally flat metric near $ \xi = 0 $ to a locally flat metric in $ \xi \geq \epsilon $. The computation leads to the following nonzero components of the Einstein tensor:
\begin{equation}
G^t_t=f''/f+\delta G,\ G^\phi_\phi=f''/f,
\end{equation}
in addition, $ G ^ i_j $ in the 2x2 block $ (i, j) = (\xi, \alpha) $ are nonzero. The expressions are lengthy and we will not explicitly write them out. The following properties were checked for them: (i) all components of the tensor $ G ^ \mu_ \nu $ vanish near $ \xi = 0 $ and in $ \xi \geq \epsilon $, where the metric is locally flat; (ii) the component $ \delta G $ after integration with the measure $(-\det g)^{1/2}d\xi $ gives an expression tending to zero with $ \epsilon \to0 $; (iii) the aforementioned $ (i, j) $-components after going to the tensor densities $ G _ {{\hat i} {\hat j}} (-\det g)^{1/2}$, where the caps denote the components in an orthonormal basis, see (2.49) \cite {Visser1996}, are finite both before and after removing $ \epsilon $-regularization, the integrals of them with respect to the coordinate volume tend to zero at $ \epsilon \to0 $.

Note that the components $ G ^ i_j $ represent the internal stresses between the layers of the tube $ 0 <\xi <\epsilon $, they depend on the chosen regularization. Physically important and independent of regularization are the external integrals of the components of the tensor $ G ^ \mu_ \nu $ over sections of constant $ t $ and $ \phi $, which gives the same answer for linear mass density and tension $ \mu = \tau = - 1/4 $, as for the straight string. For the side of the tube $ \xi = \epsilon $, the pressure forces are zero, since this surface is located entirely in the region of flat space. Thus, the radial forces arising from the tension of a curved string are compensated by the gravitational forces, in accordance with the stationary equations. The answer obtained coincides with formulas (15.17), (15.25) \cite {Visser1996}, for the dihedral solution, when choosing the geometric units ($ G = c = 1 $), for the derivation of which thin shell approximation was used. The same result was also obtained by \cite{Sorkin2,9305009} using other types of regularization.

\paragraph*{Teleporters.} In hyperbolic coordinates on the $ (r, t) $ plane, the metric can be written as
\begin{eqnarray}
&n=3,\ x=(\xi,\alpha,z),\ g=\diag(\eta,-\eta\xi^2,1),\\
&\eta=+1,\ r=\xi\cosh\alpha,\ t=\xi\sinh\alpha,\\
&\eta=-1,\ r=\xi\sinh\alpha,\ t=\xi\cosh\alpha.
\end{eqnarray}
Spacetime is represented by four maps, which correspond to four sectors on the plane $ (r, t) $. Two of them correspond to timelike vs spacelike orbits with constant $ \xi $ and variable $ \alpha $, marked by the values $ \eta = + 1 $ and $ \eta = -1 $ respectively. The other two are obtained from these by $ t \to-t $ and $ r \to-r $ reflections. Further, considering the double cover configuration of Fig.\ref{f9} for the Lorentzian case, we get a system of eight maps, shown in Fig.\ref{f12} in the center.

The key point of further construction is as follows. Usually, in general relativity a direct problem is solved, according to a given distribution of matter, a metric of spacetime is found. In this case, the expressions produced by the algorithm {\tt Einstein} are considered as a system of differential equations (Einstein field equations). These equations are solved for a given matter term with respect to the metric. This relation is nonlocal, that is, with a variation of the matter term in one patch, the metric changes in other patches, and, generally speaking, everywhere. We solve the inverse problem, for a given metric find the matter term. This calculation is performed using a series of differentiations and algebraic operations and is local. That is, when the metric is varied in one patch, the matter term changes only in this patch. As a result, we can divide the spacetime manifold into many submaps and solve our problem in each of them individually. Moreover, we can rearrange the submaps in a different order and get a different problem, with an equivalent solution. For such rearrangement, it is necessary that the metric is stitched at the borders of submaps with the required degree of smoothness. In the problem under consideration, we can provide $ C ^ \infty $-connection.

As applied to our problem, we divide the hyperbolic double cover coordinate system into many sectors with the same small $ d \alpha $. It is noteworthy that all sectors with the same $ \eta $ are equivalent to each other up to the reflections and hyperbolic rotations ($\alpha\to\alpha+Const$). Therefore, they have the same metric depending only on $ \xi $, which ensures $ C ^ \infty $-connection for any reordering of submaps within $ \eta $-sectors.

We reorder the submaps to form a four-map system, as shown in Fig.\ref{f12} on the right, as was the case with flat space. Thus, the problem will be transformed to the one that we already can solve:
\begin{equation}
n=3,\ x=(\xi,\alpha,z),\ g=\diag(\eta,-\eta f(\xi)^2,1), \label{tp3}
\end{equation}
where the regularizing function $ f (\xi) $ is introduced. If we carefully follow the permutations performed, it becomes clear that the interval $ d \alpha $ now corresponds to the double length, that is, outside the $ \epsilon $-neighborhood we have $ f (\xi) = 2 \xi $. Near zero, we have $ f (\xi) = \xi $, in order to provide a flat space there. Thus, we get the same regularizing function as in the previous subsection. It is also seen that the metric (\ref {tp3}) is related with the Euclidean case (\ref {sg3}) by a combination of Wick rotations, a formal transformation to the imaginary coordinates $ t \to iz $, $ \alpha \to i \alpha $, for $ \eta = -1 $ also $ \xi \to i \xi $. The application of the algorithm {\tt Einstein} again leads to the tensor with one nonzero component
\begin{equation}
G^z_z=f''/(\eta f),\ \eta=\pm1,
\end{equation}
related by Wick rotation with the Euclidean result (\ref {sg3G}). We keep $\eta$ in the denominator, since the obtained result is valid for all its values, not only $\pm1$. Extending this problem to the next dimension, we obtain
\begin{eqnarray}
&n=4,\ x=(\xi,\alpha,y,z),\ g=\diag(\eta,-\eta f(\xi)^2,1,1),\\
&G^y_y=G^z_z=f''/(\eta f),\ \eta=\pm1. 
\end{eqnarray}
Finally, let us roll the last two dimensions into the sphere
\begin{eqnarray}
&n=4,\ x=(\xi,\alpha,\theta,\phi),\ g=\diag(\eta,-\eta f(\xi)^2, r^2, r^2 \sin^2\theta),\\
&\eta=+1,\ r=a+(2\xi-f(\xi))\cosh\alpha+(f(\xi)-\xi)\cosh 2\alpha,\\
&\eta=-1,\ r=a+(2\xi-f(\xi))\sinh\alpha+(f(\xi)-\xi)\sinh 2\alpha,
\end{eqnarray}
where $ a $ is the radius of the sphere and interpolations are used to ensure flatness of metric at the ends of the interval $ 0 <\xi <\epsilon $. As a result, we obtain rather cumbersome expressions for the Einstein tensor, which contain a singular part
\begin{equation}
G^\theta_\theta=G^\phi_\phi=f''/(\eta f)+\delta G, 
\end{equation}
where the properties analogous to (i)-(iii) of the stargate solution are satisfied, $ \delta G $ in the integral sense is equivalent to zero, just as $ G ^ i_j $, $ (i, j) = (\xi, \alpha) $, in the appropriate normalization.

The interpretation of the result is similar to the previously considered stargate case. The Einstein tensor has two physically significant components, for stargate $ G ^ t_t = G ^ \phi_ \phi $, that is, $ - \rho = p_ \phi $; for teleporter $ G ^ \theta_ \theta = G ^ \phi_ \phi $, two components of the transverse pressure $ p_t $ on the sphere. The opposite sign of $ \rho $ and $ p_ \phi $, in which the Lorentzian metric is encoded, now takes the form of the alternating contributions $ \eta = \pm1 $ for the transverse pressure $ p_t $ on different hyperbolic maps. The numerical value of the matter term in both cases is $ p_ \phi = c_1 f '' / f $, $ p_t = \pm c_1 f '' / f $, where $ f '' / f $ gives the regularized delta function after the multiplication by $(-\det g)^{1/2}\sim f$, and the coefficient is $ c_1 = 1 / (8 \pi) $ in geometric units, about 0.03 Jupiter's mass per meter in natural units.

The main difference between the two systems is that for the teleporter the hyperbolic $ \epsilon $-neighborhood is noncompact. This manifests itself in the unboundedness of the integral $ \int d \alpha $, in contrast to the compact answer $ 2 \pi $ for the Euclidean case. Physically, for large $ \alpha $ the result has the form of two thin layers adjacent to the light cone from different sides, in which the matter term has opposite signs. In this sense, these contributions cancel each other out, like two particles with opposite masses moving in orbits tending to each other. This is an expected behavior, since the metric (\ref{morsemetr}) after removal of the regularization is flat everywhere except for the origin, also on the light cone.

Finally, let us prove that $f''(\xi)\to\delta(\xi)$ and $f''(\xi)/\eta\to0$ at $\epsilon\to0$ in the distributional sense. At first, let us specify precisely the behavior of the regularization function, shown on Fig.\ref{f10}. Let $f(\xi)$ be $C^\infty$ smooth, equal to $\xi$ at $\xi\leq b_1$, equal to $2\xi$ at $\xi\geq b_3$, with $f'(\xi)$ monotonously increasing from $1$ to $B_2$ at $b_1\leq\xi\leq b_2$, $f'(\xi)$ monotonously decreasing from $B_2$ to $2$ at $b_2\leq\xi\leq b_3$, with $0<b_1<b_2<b_3<\epsilon$ and $B_2>2$. Let $f(\xi)$ be simply rescaled with $\epsilon$, $f(\xi)\to\epsilon f(\xi/\epsilon)$, not changing the derivative $f'(\xi)$.

Consider a test function $g(\xi)$ of $C^\infty$ class with finite support on $R$, write an estimation $|g(\xi)-g(0)|\leq C|\xi|$. Evaluate $\int_0^\epsilon d\xi f''(\xi)g(0)=g(0)$ and $I =\int_0^\epsilon d\xi f''(\xi)(g(\xi)-g(0))$, $|I|\leq C\int_0^\epsilon d\xi |f''(\xi)|\xi \leq C\epsilon\int_0^\epsilon d\xi |f''(\xi)|=C\epsilon(2B_2-3)\to0$ at $\epsilon\to0$.

Consider a test function $g(x)$ of $C^\infty$ class with finite support on $R^2$, write an estimation $|g(x_1)-g(x_2)|\leq C|x_1-x_2|$. Evaluate $I = \int_0^\epsilon d\xi f''(\xi)/\eta\int_0^{+\infty} d\alpha g(x)$ over two adjacent maps. Using the symmetries, obtain $I = \int_0^\epsilon d\xi f''(\xi) \int_0^{+\infty} d\alpha (g(x_1)-g(x_2))$, where $x_1=(\xi\cosh\alpha, \xi\sinh\alpha)$, $x_2=(\xi\sinh\alpha, \xi\cosh\alpha)$, $|x_1-x_2| = \sqrt{2}\xi e^{-\alpha}$. Evaluate $|I|\leq C\int_0^\epsilon d\xi |f''(\xi)| \int_0^{+\infty} d\alpha \sqrt{2}\xi e^{-\alpha} \leq  C\sqrt{2}\epsilon \int_0^\epsilon d\xi |f''(\xi)| = C\sqrt{2}\epsilon (2B_2-3)\to0$ at $\epsilon\to0$.


\section*{References}

\begin{thebibliography}{99}
\bibitem{Visser1996}
Visser M 1996 {\it Lorentzian Wormholes: from Einstein to Hawking} (Springer)
\bibitem{1604.02082}
Lobo F S N 2016 From the Flamm-Einstein-Rosen bridge to the modern renaissance of traversable wormholes {\it Int. J. Mod. Phys. D} {\bf 25} 1630017 (arXiv:1604.02082)
\bibitem{Geroch1}
Geroch R P 1967 Singularities in the spacetime of general relativity: Their definition, existence, and local characterization, PhD thesis (Princeton University)
\bibitem{Geroch2}
Geroch R P 1967 Topology in general relativity {\it J. Math. Phys.} {\bf 8} 782-6
\bibitem{Geroch-Horowitz}
Geroch R P and Horowitz G T 1979 Global structure of spacetimes, in Hawking S W  and Israel W, eds. {\it General Relativity: An Einstein Centenary Survey} (Cambridge University Press) pp 212-93
\bibitem{Borde}
Borde A 1994 Topology change in classical general relativity (arXiv:gr-qc/9406053)
\bibitem{Hawking1}
Hawking S W 1992 The chronology protection conjecture, in Sato H, ed. {\it Proceedings of the 6th Marcel Grossmann Meeting, Kyoto, Japan, June 24-29, 1991} (World Scientific) pp 3-13
\bibitem{Hawking2}
Hawking S W 1992 Chronology protection conjecture {\it Phys. Rev. D} {\bf 46} 603-11
\bibitem{Hawking-Ellis}
Hawking S W and Ellis G F R 1973 {\it The Large Scale Structure of Spacetime} (Cambridge University Press)
\bibitem{Yodzis1972}
Yodzis P 1972 Lorentz cobordism {\it Commun. Math. Phys.} {\bf 26} 39-52
\bibitem{Sorkin1}
Sorkin R D 1986 Topology change and monopole creation {\it Phys. Rev. D} {\bf 33} 978
\bibitem{Sorkin2}
Louko J and Sorkin R D 1997 Complex actions in two-dimensional topology change {\it Class. Quant. Grav.} {\bf 14} 179-204 (arXiv:gr-qc/9511023)
\bibitem{Horowitz1991}
Horowitz G 1991 Topology change in classical and quantum gravity {\it Class. Quant. Grav.} {\bf 8} 587
\bibitem{9109030}
Horowitz G 1991 Topology change in general relativity {\it Proc. of the Sixth Marcel Grossmann Meeting, Kyoto, Japan, June 24-29} (arXiv:hep-th/9109030)
\bibitem{Morris-etal}
Morris M S, Thorne K S and Yurtsever U 1988 Wormholes, time machines, and the weak energy condition {\it Phys. Rev. Lett.} {\bf 61} 1446-9
\bibitem{9305009}
Balasin H and Nachbagauer H 1993 What curves the Schwarzschild geometry {\it Class. Quant. Grav.} {\bf 10} 2271-8 (arXiv:gr-qc/9305009)
\bibitem{Raju1982}
Raju C K 1982 Junction conditions in general relativity {\it J. Phys. A} {\bf 15} 1785-97
\bibitem{CosmicCens}
Joshi P S and Saraykar R V 1987 Cosmic censorship and topology change in general relativity {\it Phys. Lett. A} {\bf 120} 111-4
\bibitem{vickers1}
Vickers J A G 1990 Quasi-regular singularities and cosmic strings {\it Class. Quant. Grav.} {\bf 7} 731-41
\bibitem{9907105}
Vickers J A G and Wilson J P 2000 Generalized hyperbolicity in conical spacetimes {\it Class. Quant. Grav.} {\bf 17} 1333-60 (arXiv:gr-qc/9907105)
\bibitem{9605060}
Clarke C J S, Vickers J A and Wilson J P 1996 Generalized functions and distributional curvature of cosmic strings {\it Class. Quant. Grav.} {\bf 13} 2485-98 (arXiv:gr-qc/9605060)
\bibitem{9711069}
Ionicioiu R 1997 Building blocks for topology change in 3D {\it Report DAMTP-97-127} (arXiv:gr-qc/9711069)
\bibitem{0410087}
Lobo F S N 2005 Energy conditions, traversable wormholes and dust shells {\it Gen. Rel. Grav.} {\bf 37} 2023-38 (arXiv:gr-qc/0410087)
\bibitem{0505150}
McCabe G 2005 The topology of branching universes {\it Found. Phys. Lett.} {\bf 18} 665-76 (arXiv:gr-qc/0505150)
\bibitem{conical_spacetimes}
Hörmann G 2015 Conical spacetimes and global hyperbolicity {\it Novi Sad J. Math.} {\bf 45} 215-29 (arXiv:1501.00672)
\bibitem{0610441}
Kardashev N S, Novikov I D and Shatskiy A A 2007 Astrophysics of wormholes {\it Int. J. Mod. Phys. D} {\bf 16} 909-26 (arXiv:astro-ph/0610441)
\bibitem{qtp2017}
Ren J G et al 2017 Ground-to-satellite quantum teleportation {\it Nature} {\bf 549} 70-73 (arXiv:1707.00934)
\bibitem{Visser1}
Visser M 1989 Traversable wormholes from surgically modified Schwarzschild spacetimes {\it Nucl. Phys. B} {\bf 328} 203-12 (arXiv:0809.0927)
\bibitem{Visser2}
Visser M 1989 Traversable wormholes: some simple examples {\it Phys. Rev. D} {\bf 39} 3182-4 (arXiv:0809.0907)
\bibitem{Blau2018}
Blau M 2018 {\it Lecture Notes on General Relativity} (University of Bern)
\bibitem{math-codes}
Hartle J B 2003 {\it Gravity: An Introduction to Einstein's General Relativity} (Addison-Wesley)
\bibitem{1407.6026}
Battarra L, Lavrelashvili G and Lehners J L 2014 Creation of wormholes by quantum tunneling in modified gravity theories {\it Phys. Rev. D} {\bf 90} 124015 (arXiv:1407.6026)
\bibitem{GuthBubble}
Waldrop M M 1987 Do-it-yourself universes {\it Science} {\bf 235} 845
\bibitem{EuclideanWormholes}
Hebecker A, Mikhail T and Soler P 2018 Euclidean wormholes, baby universes, and their impact on particle physics and cosmology {\it Front. Astron. Space Sci.} {\bf 5} 35 (arXiv:1807.00824)
\bibitem{0205066}
Barceló C and Visser M 2002 Twilight for the energy conditions? {\it Int. J. Mod. Phys. D} {\bf 11} 1553-60 (arXiv:gr-qc/0205066)
\bibitem{0602086}
Ashtekar A, Pawlowski T and Singh P 2006 Quantum nature of the Big Bang {\it Phys. Rev. Lett. } {\bf 96} 141301 (arXiv:gr-qc/0602086)
\bibitem{0604013}
Ashtekar A, Pawlowski T and Singh P 2006 Quantum nature of the Big Bang: an analytical and numerical investigation {\it Phys. Rev. D} {\bf 73} 124038 (arXiv:gr-qc/0604013)
\bibitem{0607039}
Ashtekar A, Pawlowski T and Singh P 2006 Quantum nature of the Big Bang: improved dynamics {\it Phys. Rev. D} {\bf 74} 084003 (arXiv:gr-qc/0607039)
\bibitem{1401.6562}
Rovelli C and Vidotto F 2014 Planck stars {\it Int. J. Mod. Phys. D} {\bf 23} 1442026 (arXiv:1401.6562)
\bibitem{1409.1501}
Barceló C, Carballo-Rubio R, Garay L J and Jannes G 2015 The lifetime problem of evaporating black holes: mutiny or resignation {\it Class. Quant. Grav. } {\bf 32} 035012 (arXiv:1409.1501)
\bibitem{1701.01569} 
Klimenko S V, Nikitin I N and Nikitina L D 2017 Numerical solutions of Einstein field equations with radial dark matter {\it Int. J. Mod. Phys. C} {\bf 28} 1750096 (arXiv:1701.01569)
\bibitem{0411062} 
Hamilton A J S and Pollack S E 2005 Inside charged black holes: II. Baryons plus dark matter {\it Phys. Rev. D} {\bf 71} 084032 (arXiv:gr-qc/0411062)
\bibitem{0808.2870}
Ghez A M et al 2008 Measuring distance and properties of the Milky Way's central supermassive black hole with stellar orbits {\it Astrophys. J.} {\bf 689} 1044-62 (arXiv:0808.2870)
\end{thebibliography}
\end{document}